\documentclass[fleqn,usenatbib]{mnras}

\usepackage{newtxtext,newtxmath}

\usepackage[T1]{fontenc}

\DeclareRobustCommand{\VAN}[3]{#2}
\let\VANthebibliography\thebibliography
\def\thebibliography{\DeclareRobustCommand{\VAN}[3]{##3}\VANthebibliography}

\usepackage{graphicx}	
\usepackage{amsmath}	
\usepackage{gensymb}
\usepackage{hyperref}
\usepackage{ulem}

\title[An energy approach to pulsar-disc interaction]{An energy approach to pulsar-disc interaction: disc stability and implications for transitional millisecond pulsars}

\author[Vurgun et al.]{
Eda Vurgun,$^{1}$\thanks{E-mail: eda.vurgun@upc.edu}\thanks{edavrgn@gmail.com}
Domingo García-Senz,$^{1,2}$\thanks{E-mail: domingo.garcia@upc.edu}
Manuel Linares$^{3,1}$
and K.~Yavuz Ek{\c s}i$^{4}$
\\
$^{1}$Departament de Física, EEBE, Universitat Politècnica de Catalunya, c/Eduard Maristany 16, E-08019 Barcelona, Spain\\
$^{2}$Institut d’Estudis Espacials de Catalunya, Gran Capità 2-4, E08034 Barcelona, Spain\\
$^{3}$Department of Physics, Norwegian University of Science and Technology, NO-7491 Trondheim, Norway\\
$^{4}$Istanbul Technical University, Faculty of Science and Letters, Physics Engineering Department, 34469 Istanbul, Turkey
}

\date{Accepted XXX. Received YYY; in original form ZZZ}

\pubyear{2015}

\begin{document}
\label{firstpage}
\pagerange{\pageref{firstpage}--\pageref{lastpage}}
\maketitle

\begin{abstract}
 The stability of an accretion disc surrounding a millisecond pulsar is analysed from an energetic point of view, using magnetohydrodynamic simulations that consider realistic disc structures and a variety of magnetic field inclination angles. The time-averaged components of the magnetic field interact with the disc through ohmic dissipation, which causes heating and partial evaporation of its innermost region. The stability of the disc right after the magnetic field is turned on is analysed as a function of the location of the inner radius of the disc and the magnetic inclination angle. Our results show that the disc is severely altered in those cases where its inner radius lies well beyond the light cylinder and the magnetic axis is not totally aligned with the neutron star spin axis. Overall, the results of the simulations agree with those obtained in previous works where analytical or semi-analytical energy models were also used to discuss the stability of the disc. The implications for the understanding of the transitional millisecond pulsars are discussed. We briefly mention implications of our results for low-mass X-ray binaries and supernova fallback discs.
\end{abstract}

\begin{keywords}
neutron stars -- accretion discs -- magnetohydrodynamics
\end{keywords}



\section{Introduction}
\label{sec:intro}

  Neutron stars (NSs) in low-mass X-ray binaries (LMXBs) are thought to be spun up to millisecond rotational periods by accreting matter transferred from a companion star and evolve into millisecond pulsars \citep[MSPs;][]{alpar1982}. When the rate of mass transfer decreases at the end of the accretion phase (X-ray bright), a radio MSP turns on, powered by the loss of rotational energy of the NS. A subclass known as compact binary or ``spider'' MSPs has emerged over the past decade, with low-mass (redbacks) or ultralight (black widows) companion stars in compact orbits (periods of about a day or shorter, and with orbital separations $a \sim 10^{11}$~cm).
  
  Occasionally, some redback MSPs experience transitions from a rotation-powered ``pulsar state'' to an accretion ``disc state'', and vice versa \citep{2014Linares,2022Papitto}.
  They are known as transitional millisecond pulsars (tMSPs).
  In the disc state, tMSPs show broad double-peaked optical emission lines consistent with the presence of an accretion disc and an enhanced X-ray luminosity $L_{\rm X} \sim 10^{33}-10^{34}$~erg~s$^{-1}$ (still 2-3 orders of magnitude lower than the $L_{\rm X}$ of LMXBs in their active outburst states).
  In the pulsar state, tMSPs emission is dominated by the companion star in the optical band and $L_{\rm X}$ is lowest at $\sim 10^{31}-10^{32}$~erg~s$^{-1}$.
  All evidence for an accretion disc vanishes, indicating that tMSPs manage to fully eject or evaporate their discs in the state transitions.
  Such transitions occur on timescales of a few days, and the disc and pulsar states can persist for at least years or decades \citep[e.g.,][]{stappers2014}.

From a theoretical point of view, the details of the physical mechanisms driving the changes in the state of tMSPs are not yet well understood, while the different states are roughly associated with different modes of disc-magnetosphere interaction \citep[see][for review]{1992Lipunov, Rom15,abo+24}. Starting with the work by \cite{shvartsman1970} several authors identified the different modes of interaction between the magnetosphere of the neutron star and the surrounding accretion flow with analytical approaches, albeit phenomenological \citep{shvartsman1971, pringle1972, davidson1973}. According to these works, the inner boundary of the quasi-spherical accretion flow is at the Alfvén radius,
\begin{equation}
    R_{\rm A} = \left( \frac{\mu^2}{\sqrt{2GM_{\rm NS}} \dot{M}}\right)^{2/7}\,,
    \label{eq:Alfven}
\end{equation}
where the ram pressure (twice the kinetic energy density) of the accreting matter is equal to the stellar magnetic pressure (the same as the magnetic energy density) and with $\mu$, $M_{\rm NS}$ and $\dot{M}$ being the magnetic dipole moment, the mass of the NS and the mass-accretion rate, respectively. In the case of a thin disc that interacts with the magnetosphere of the star, one needs to compare the material stresses in the disc with the magnetic stresses, rather than the pressures, but finds a similar scaling $R_{\rm in} = \beta R_{\rm A}$ \citep[see e.g.,][]{gho79a,gho79b, Wang96}.
The steady accretion of matter onto the compact object is possible if the star is not rotating more rapidly than the Keplerian speed at the inner boundary of the accretion flow. This so-called \textit{accretion regime} corresponds to what we observe as accreting X-ray pulsars.
Should the mass flow rate be lower, the inner radius of the disc, $R_{\rm in}$, would move beyond the corotation radius, 
\begin{equation}
R_{\rm co} = \left(\frac{GM_{\rm NS} P^2}{4\pi^2}\right)^{1/3}\,,
\label{eq:Corot}
\end{equation}
where $P$ is the rotation period of the neutron star. In this case,  the matter would encounter a centrifugal barrier which is the so-called \textit{propeller} regime \citep{shvartsman1971,1975Illarionov}. At this stage, accretion can not proceed steadily and strong outflows may occur if $R_{\rm in} \gg R_{\rm co}$ \citep{lov+99, Rom+03pr,Rom+04, Rom+18}. If the mass flow rate is even lower, the neutron star could act as a radio pulsar (ejector stage). For this to happen, it is presumed that the inner disc is beyond the light cylinder radius, 
\begin{equation}
R_{\rm LC} = \frac{c P}{2\pi}~,
\label{eq:LC}
\end{equation}
or the disc is totally evaporated by ablation of the central object.

Direct numerical simulations of the interaction between the magnetic field of the central star and the accretion disc were carried out in connection with T-Tauri stars \citep[e.g.,][]{Hay+96, zanni2009, zanni2013, tak+18} and in the context of compact binary systems \citep[e.g.,][]{miller1997, Rom+02, Rom+03inc, Rom+08, parfrey2017, cik+22, par24}. These works focused on the study of the torque exerted by the disc onto the central star by the magnetosphere and were carried out in axial symmetry with a general relativistic (GR) MHD code in the latter case. An axisymmetric GR-MHD code was also used by \cite{parfrey2017b} to numerically reproduce, for the first time, the different possible disc-magnetosphere interaction regimes in these compact binary sources. However, that study focused on a narrow range of radial distances from the NS, close to the radius of the light cylinder at $R_{\rm LC}$, using a torus as a mass reservoir instead of a realistic accretion disc. It is also worth noting that all these studies above assumed that the structure of the magnetic field around the NS can be described with the dipole-like geometry characteristic of an aligned magnetic rotator. Accretion onto oblique \citep{romanova2021,mur+24,das24} and non-dipolar \citet{das+22} fields was also considered.

The interaction of the pulsar wind with the secondary star wind and its connection with the tMSP problem has recently been outlined by \cite{guerra2024} with 2D-Cartesian hydrodynamic simulations. Promising full three-dimensional MHD simulations of the structure of the wind around the pulsar have recently been developed by \cite{cerutti2016,cerutti2020} with a particle-in-cell (PIC) code. Because these calculations are free of geometrical constraints, they can completely incorporate pulsar radiation emitted by oblique rotators. 

\cite{eksi05} investigated the different outcomes of the pulsar-disc interaction by comparing the energy density of the electromagnetic fields, as given by the vacuum solution of \cite{deutsch55}, with the kinetic energy density of the accretion flow. The \cite{deutsch55} solution \citep[see also][]{michel99} is a suite of analytical expressions of the time-dependent electric ($\mathbf {E}$) and magnetic ($\mathbf {B}$) fields produced by an oblique magnetic rotator, a sphere with infinite conductivity, at any distance from its centre.  To carry out the comparisons, \cite{eksi05} made useful analytical approximations to the electromagnetic energy density $(\langle E^2\rangle + \langle B^2 \rangle)/(8\pi)$, where the brackets indicate a period-averaged quantity. In this way, they managed to explore the different outcomes of the pulsar-disc electromagnetic interaction on purely energetic grounds. It is well understood, since the seminal work of \citet{gol69}, that a real pulsar would have a corotating plasma which would lead to an electromagnetic field distribution rather different from the Deutsch solution, and that the presence of the disc would also change the structure dramatically. In the absence of a suitable analytical solution that is continuous across the light cylinder separating the near- and radiation-zone, the Deutsch solution is still a reasonable choice for investigating the disc magnetosphere interaction near the light cylinder radius.

In this work, we use the Deutsch solution to study the interaction of the magnetic field with the nearby accretion disc using energy arguments, similar to those described in \cite{eksi05}, but carrying out detailed MHD numerical simulations which incorporate realistic disc structures. Particular emphasis is placed on scenarios where the inner disc (truncated at a radial distance $R_{\rm in}$) is quite far from the NS, barely investigated in the current tMSP literature. To perform the simulations, we use an axisymmetric MHD smoothed particle hydrodynamics code (Axis-SPHMHD) recently developed \citep{gsenz_23}. To our knowledge, this is the first time an SPHMHD code has been used to simulate MSP state transitions.  Our study focusses on the interaction of the averaged magnetic field of the NS, $\langle B^2 \rangle^{1/2}$, with the disc, which in most cases is severely injured by the heating due to disc-surface induced electric currents when the pulsar's magnetic field is activated, especially for oblique rotators. This is perhaps the simplest setting to get insight into the stability and endurance of the disc in the presence of the NS magnetic field.  In our  simulations the disc's inner radius\footnote{Here taken as the closest point of the disc from the NS where the radial velocity is still small, $v_r \ll v_{\rm orb}$}  and the magnetic inclination have been conveniently varied so that the ensuing impact on the disc structure is further analysed.

Section \ref{sec:B-disk-interaction} gives information on the main features of the electromagnetic power radiated by a rotating NS and the expected effect of that power on the accretion rate from the disc onto the NS. Sections \ref{sec:mhd} and \ref{sec:Initial models} summarise the main features of the MHD method used to perform the simulations and the procedure to obtain the initial disc configurations in equilibrium. The main results of the simulations are presented and discussed in Sects.~\ref{sec:calculations} and \ref{sec:discussion} to finally summarise and highlight the main findings of this work and comment on future developments in the conclusions section, Sect. \ref{sec:conclusions}.    

\section{Electromagnetic pulsar-disc interaction near and beyond the transition region}

\label{sec:B-disk-interaction}

The effects of the electromagnetic radiation from the pulsar onto the nearby accretion disc is a forefront topic of great complexity. It involves many pieces of physics such as the structure of the time-dependent magnetic $\mathbf {B}(t)$ and electric fields $\mathbf{E}(t)$ radiated by the pulsar and its further interaction with the surrounding plasma. The electromagnetic radiation may interact directly with the surrounding matter, as in \cite{eksi05}, or the interaction with the disc is mediated by the electromagnetic pulsar wind \citep{parfrey2017} with some contribution of a relativistic particle wind \citep{veledina2019}. However, the qualitative behaviour of the pulsar-disc interaction and the connection with the tMSP phenomena, through the appearance-disappearance of the accretion disc, can be understood with a simple piecewise defined energy density model, as shown by \cite{1975Illarionov, campana1998, papitto2022}.

\begin{figure*}
\centering
\includegraphics[scale=0.31]{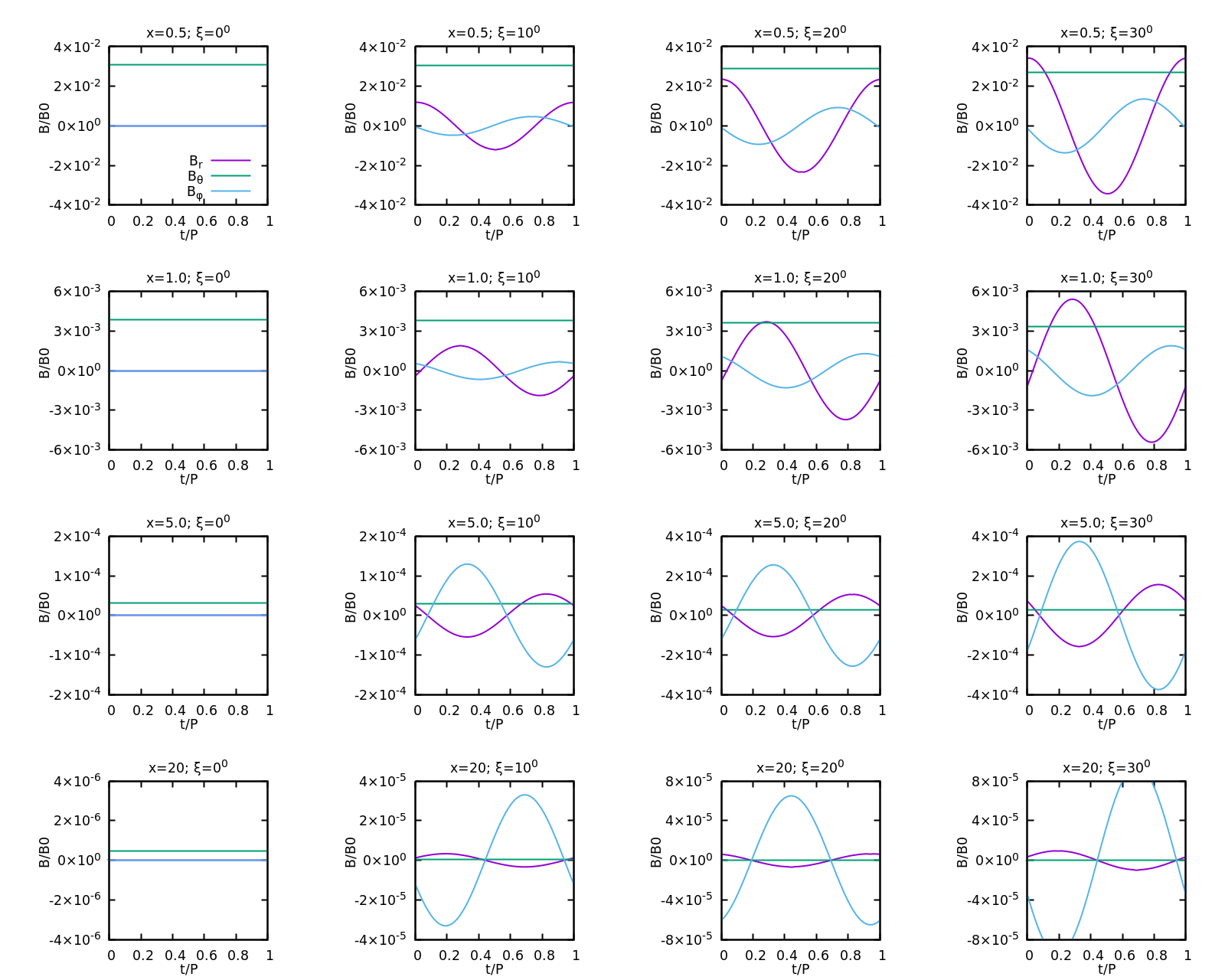}
    \caption{Components of the magnetic field vector (in units of $B_0$, the magnetic field strength at the neutron star equator) for tilt angles, from left to right,  $\xi=0\degree$ (aligned rotator) and $\xi=10\degree, 20\degree, 30\degree$ (oblique rotators) as function of normalized time ($P$ is the spin period of the pulsar) at four  distances from the pulsar (from top to bottom and in units of the light cylinder radius) $x=r/R_{\rm LC}= 0.5, 1, 5, 20$ respectively.}
    \label{deustch_1}
\end{figure*}

\begin{figure*}
\centering
\includegraphics[scale=0.315]{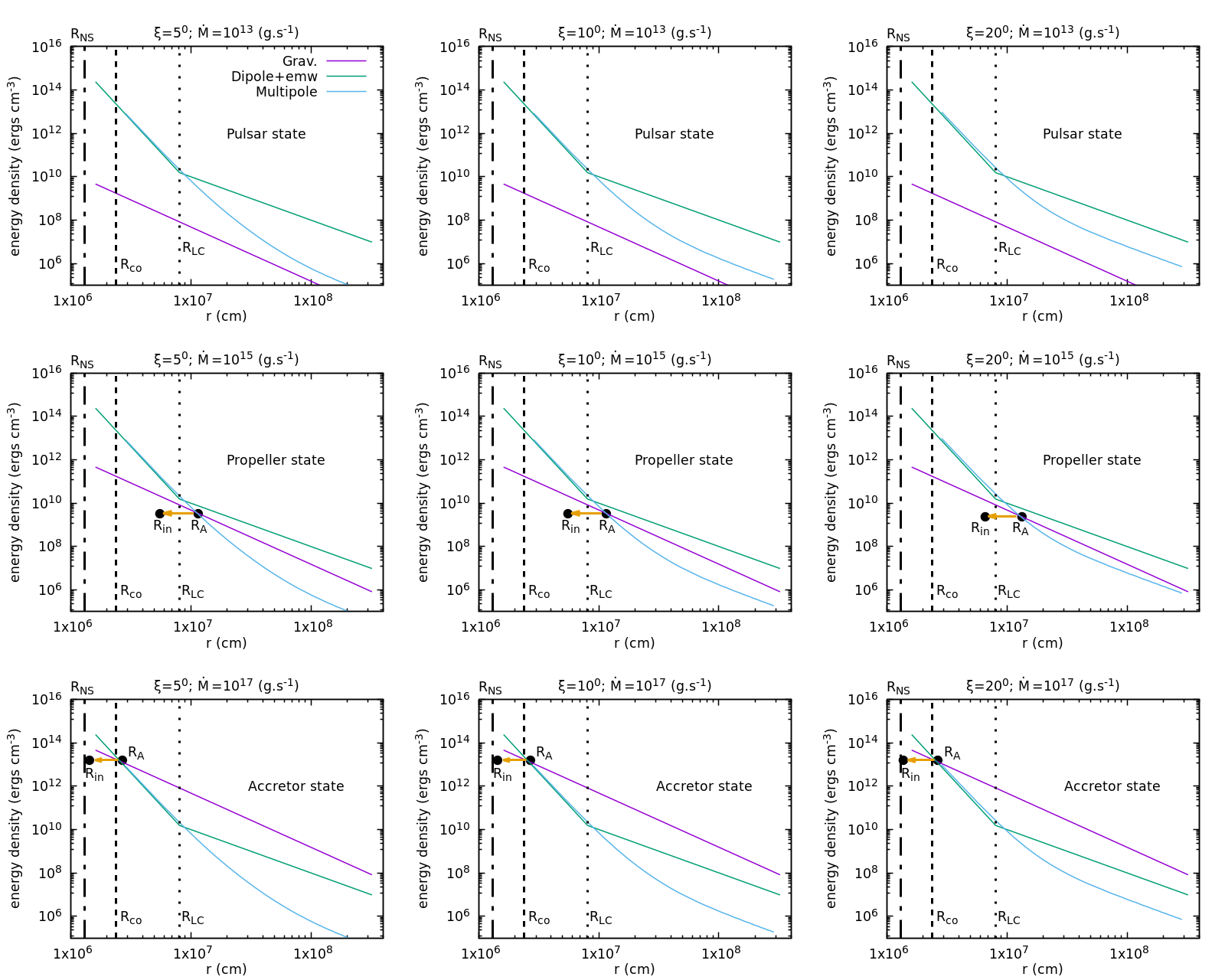}
    \caption{Accretion versus magnetic energy density for different accretion rates, $\dot {M}$, and tilt angles, $\xi$, as a function of the distance to the neutron star. The head of the yellow arrow points to the accretion radius taken as one-half of the Alfvén radius $R_{\rm A}$. The vertical lines indicate the neutron star radius, the co-rotation radius, and the light cylinder radius respectively.}
    \label{power_1}
\end{figure*}

Consider first that the energy radiated by the pulsar is that of a magnetised oblique rotator in a vacuum, whose properties have been thoroughly described by \cite{deutsch55} and \cite{michel99}. A glimpse of Fig.\ref{deustch_1} gives an idea of the richness of $\mathbf {B}(t)$ around the spinning neutron star. The components of ${\mathbf B}(t)$ in spherical coordinates fluctuate from pure dipole ($B_\theta\neq 0; B_r=B_\varphi=0$) for strictly aligned rotators (first column) to azimuthal, $B_\varphi$-dominated radiation for misaligned rotators at large distances. The latter case corresponds to a typical electromagnetic wave (e.m.w.) propagating through vacuum. It is worth noting that the aligned rotator solution is unstable and is not very realistic, indeed because even a small value of the magnetic axis angle produces a big change in the electromagnetic field around \citep{eksi05}. Still, the constant field $B_\theta$ takes over the other oscillating components for $x \equiv r/R_{\rm LC}\le 1$ and $\xi \le 20\degree$,  where $\xi$ is the inclination of the magnetic axis, as shown in the first two rows in Fig.~\ref{deustch_1}.

The accretion of matter, which is ultimately responsible for the X-ray state of the NS (XMSP), is only plausible if the electromagnetic pressure remains below the ram pressure of the infalling matter coming from the disc. Figure~\ref{power_1} gives the value of these pressures, or energy densities ($u_x$) where $x$ is the associated physical magnitude, as a function of the distance to the NS and for different accretion rates and magnetic angles \citep[e.g.,][and references therein]{papitto2022}.  In particular, the green line in Fig.\ref{power_1} shows the approximate electromagnetic energy density, 
\begin{equation}
  u_{\rm em}=
\begin{cases}
\displaystyle{ \frac{\mu^2}{8\pi r^6}}; & \text{dipolar for $r \ll R_{\rm LC}$,}\\ 
\displaystyle{ \frac{\mu^2}{8\pi R_{\rm LC}^4 r^2}};& \text{e.m.w. for $r \gg R_{\rm LC}$,}\\
\end{cases}
\label{eq:power_rad}
\end{equation}
 and the magenta line is that of the gravitational density of energy of the infalling gas,
\begin{equation}
    u_{\rm grav}=\frac{\sqrt{2GM_{\rm NS}} ~\dot {M}}{4\pi r^{5/2}},\\
    \label{eq:power_grav}
\end{equation}
  The intersection of lines defined by Eqs.~\ref{eq:power_rad} and \ref{eq:power_grav} roughly defines the Alfvén radius $R_{\rm A}$, which sets the inner radius of the disc, usually taken as $R_{\rm in}=\beta R_{\rm A}$ ($\beta\simeq 0.5$). Depending on the locus of $R_{\rm in}$, different states of the tMSP occur.   

In order for a stable disc-magnetosphere configuration to be possible, the electromagnetic energy density should be increasing more steeply than the gravitational potential energy as $r$ becomes smaller (so that moving inward from the equilibrium position, $R_{\rm in}$, by $\delta R_{\rm in}$, the stronger pressure at $R_{\rm in}-\delta R_{\rm in}$ would restore the disc back. In fact, this condition is well satisfied for $r\le R_{\rm LC}$ where $u_{\rm em} \propto r^{-6}$ while $u_{\rm grav} \propto r^{-5/2}$. However, this is not satisfied in the radiation zone $r>R_{\rm LC}$ since $u_{\rm em}\propto r^{-2}$ in this region. Yet, there is actually a continuous transition from the near-zone static fields to the far-zone radiation fields across $R_{\rm LC}$ where the slope of the energy density changes from $r^{-6}$ to $r^{-2}$ as found by \citet{eksi05} from the Deutsch solutions
\begin{equation}
u_{\rm em}=\frac{\mu ^{2}}{8\pi R_{\rm LC}^{6}}x^{-6}\left[ \left( 1+\frac{4}{9}x^{2}\right)
\cos ^{2}\xi +\left( x^{4}+2x^{2}+\frac{5}{2} \right) \sin ^{2}\xi \right]
\label{eq:u_tot2}
\end{equation}
where $x \equiv r/R_{\rm LC}$.
This allows a stable disc-magnetosphere interaction even beyond the light cylinder \citep{eksi05}.

Therefore, the rough estimation above, Eq.~\ref{eq:power_rad}, can be made more precise if the whole multipolar solution of the radiating fields is considered, as shown by the blue line in Fig.~\ref{power_1}. This line defines the total electromagnetic energy $(\langle E^2 \rangle + \langle B^2 \rangle)/(8\pi)$. The time averages were calculated numerically with the complete expressions of the Deutsch fields, as detailed in \cite{michel99}. Our results are similar to those obtained by \cite{eksi05} using analytical approaches (Eq.~\ref{eq:u_tot2}), but our numerical  
approach can handle polar angles $\theta\ne \pi/2$ and is not necessarily restricted to very small $R_{\rm NS}/R_{\rm LC}$ ratios. 
The intersection of the blue and magenta lines in Fig.~\ref{power_1} is a better approach to the radius $R_{\rm A}$ and $R_{\rm in}$ than the usual estimates based on the dipolar field strength \citep[e.g.,][]{campana1998}. According to Fig.~\ref{power_1} and depending on the accretion rate, here arbitrarily chosen for illustrative purposes, we can distinguish three tMSP states:   

a) the pulsar state, calculated with $\dot M=10^{13}$ g~s$^{-1}$ (no intersection, first row in Fig.~\ref{power_1});

b) the propeller state, calculated with $\dot M=10^{15}$~g~s$^{-1}$ (second row in Fig.~\ref{power_1}) and leading to $R_{\rm co}\le R_{\rm in}\le R_{\rm LC}$, where the co-rotation radius, $R_{\rm co}$, is given by Eq.~\ref{eq:Corot}, and 
%
%

 c) the outburst, or ``accretor'' state for higher accretion rates $\dot M> 10^{17}$ g~s$^{-1}$ (third row in Fig.~\ref{power_1}), where the accretion of matter is not blocked and $R_{\rm in}\le R_{\rm co}$.

Figure \ref{power_1} also shows that the slope of the full Deutsch solution agrees with the pure dipole for $r\le R_{\rm LC}$ but diverges from the line defined by Eq.~\ref{eq:power_rad} for $r > R_{\rm LC}$. Nevertheless, at larger distances, the Deutsch and e.m.w lines go parallel, especially at magnetic angles $\xi\ge 20\degree$. 

It is worth noting that the slope of the gravitational energy density  $u_{\rm grav}\propto r^{-5/2}$ (Eq.~\ref{eq:power_grav}) is that of the free-fall  material in the accretion column, which differs from the density of energy stored in the rotational velocity field in the disc,\footnote{Assuming Keplerian rotating velocity $v_\varphi= \sqrt{GM_{\rm NS}/r}$ and a density profile typical of the thin disc approach $\rho\propto \dot M_{\rm in} r^{-15/8}$ \citep{frank02}.} $u_{\rm rot}\simeq \rho v_{\varphi}^2 \propto r^{-23/8}$. The intersection between the electromagnetic energy density and $u_{\rm rot}\propto r^{-23/8}$ better defines the truncation radius of the disc, $R_{\rm T}$, which would settle to the right of $R_{\rm A}$ in Fig.~\ref{power_1}. Since this scaling is even steeper than that of $u_{\rm grav}$, the region where the disc-magnetosphere configuration is stable would be more limited.

In this work, we still use the energy approach to the pulsar-disc interaction as in \cite{eksi05}, and we take a step forward by introducing a realistic disc structure in the simulation. This allows us to study the interaction of the multipolar radiation on different regions of the accretion disc in more detail. To make the study numerically tractable, we adopt two simplifying hypotheses: 

\begin{itemize} 
\item Following \cite{eksi05} the time-dependent magnetic field is replaced by its time-averaged value on a period $P$ so that $\mathbf B$ depends only on the spatial coordinates $\mathbf B(r) = \langle B^2_r \rangle^{1/2} \hat r + \langle B^2_\theta \rangle^{1/2} \hat \theta + \langle B^2_\varphi \rangle^{1/2} \hat \varphi$, where $(\hat r, \hat \theta, \hat \varphi)$ are unit vectors in spherical coordinates and $\langle B_i^2 \rangle(r) =1/P \int_0^P B_i^2(r)$. This is justified because the period of the pulsar is so small that the disc reacts to the average field of many pulsar rotations.     

\item We consider that the disc has axial symmetry with respect to the NS axis of rotation. The axisymmetric approach assumes that any magnitude along a particular longitudinal section of the disc at a distance $s$ from the rotation axis has, after being time-averaged, the same behaviour as that at the same distance in any other section and can therefore be described with only two variables $(s,z)$ where $s=\sqrt{x^2+y^2}$. Working in axial geometry boosts the resolution and allows us to simulate a larger fraction of the accretion disc investing moderate computational resources.  
\end{itemize}

\section{Axial MHD simulation code and setting of the initial configuration}

\label{sec:mhd}

The simulations were carried out with the Newtonian axisymmetric code Axis-SPHNYX recently developed by \cite{gsenz_23} to handle scenarios with axial symmetry and magnetic fields using the smoothed particle hydrodynamics technique. Although general relativistic effects can be important for millisecond pulsars in the accretion regime \citep[see e.g.][]{psa99, parfrey2017b} where the inner disc is closest to the star, they are likely to become less prominent in the regime we investigate where the inner radius is closer to the light cylinder radius. Thus, a Newtonian approach is accurate enough for the goals of this work. The introduction of special relativistic effects is beyond the scope of this work and is left for future extensions of the SPMHD code \citep[see, e.g.,][]{Rosswog2010}. The axial MHD code relies on the 3D-SPMHD code by \cite{pri12, price18} but restricts it to the axisymmetric $(r,z)$ plane ($r=\sqrt{x^2+y^2}$) and adding the pertinent non-Cartesian contributions (called the hoop-stress terms)  to the hydrodynamic Euler equations. Because of its Lagrangian nature, it is adequate to describe systems with complicated geometries surrounded by large voids, such as accretion discs immersed in the external magnetic field created by the neutron star. Assuming axial symmetry allows for a good spatial resolution, which is a crucial requirement for studying the evolution of thin accretion discs characterised by a low height/radius ratio. The typical ratio between the height, $H$, and the radial size, $\Delta r$, of the disc in our simulations is $H/\Delta r \simeq 0.03$. For similar resolution, working in full 3D is, of course, a better option since the reduction to axial symmetry is, in some ways, an ad hoc artificial constraint and hydromagnetic instabilities are better represented in 3D. Promising computational advances are making Exascale calculations with SPH affordable \citep{Cavelan2020} and accretion disc simulations could benefit from that.

For completeness, we summarise below the main axial SPMHD equations used in this work: 

\begin{itemize}

\item {\sl Mass conservation}
    
\begin{equation}
\eta_a=\sum_{b=1}^{n_b} \varepsilon_b~m_b W_{ab}(h_a)
\label{eq:densitymhd}
\end{equation}

\item {\sl Momentum equations}

\begin{equation}
\begin{split}
a^r_a=~ & 2\pi\frac{\left(p_a+\frac{B_a^2}{2\mu_0}-\frac{(B_a^\varphi)^2}{\mu_0}\right)}{\eta_a}+\\
&2\pi \sum_{b=1}^{n_b}  m_b \left(\frac{S_a^{ri} \vert r_a\vert}{\eta_a\eta_b}{\mathcal A}_{ab}^{i}(h_a)+
 \varepsilon_b\frac{S_b^{ri} \vert r_b\vert}{\eta_a\eta_b} {\mathcal A}_{ab}^{i}(h_b)\right)\,, 
\end{split}
\label{eq:mhdaccel_r}
\end{equation}

\begin{equation}
a^z_a=  2\pi\sum_{b=1}^{n_b} m_b\left(\frac{S_a^{zi} \vert r_a\vert}{\eta_a\eta_b}{\mathcal A}_{ab}^{i}(h_a)+\varepsilon_b\frac{S_b^{zi} \vert r_b\vert}{\eta_a\eta_b}{\mathcal A}_{ab}^{i}(h_b)\right)\,,
\label{eq:mhdaccel_z}
\end{equation}

\begin{equation}
\begin{split}
a^\varphi_a&=  2\pi\left(\frac{B_a^r B_a^{\varphi}}{\mu_0\eta_a}\right)+\\
&2\pi \sum_{b=1}^{n_b}  m_b \left(\frac{S_a^{\varphi i} \vert r_a\vert}{\eta_a\eta_b}{\mathcal A}_{ab}^{i}(h_a)+
 \varepsilon_b\frac{S_b^{\varphi i} \vert r_b\vert}{\eta_a\eta_b} {\mathcal A}_{ab}^{i}(h_b)\right)\,,
\end{split}
\label{eq:mhdaccel_phi}
\end{equation}
where $h_a, h_b$ are the smoothing lengths of particles $a$, $b$ respectively (i.e. the local resolution), $m_a$ and is the mass of the SPH particle so that $\sum_a m_a = M_{\rm disc}$. Repeated indices in $\{i=r,z\}$~are summed.

\item {\sl Energy equation}

\begin{equation}
\begin{split}
\frac{du_a}{dt}=-2\pi\frac{p_a}{\eta_a} v_{r_a}+
&2\pi\frac{p_a \vert r_a\vert }{\eta_a}\sum_{b=1}^{n_b} \frac{m_b}{\eta_b} \left( v^i_{ab}~{{\bf\mathcal A}^i_{ab}}(h_a)\right)\,.
\label{eq:energy}
\end{split}
\end{equation}
where $S_a^{ij}$ are the components of the stress tensor of particle $a$,
\begin{equation}
    S_a^{ij}=-\left(p_a+\frac{1}{2\mu_0}B_a^2\right)\delta^{ij}+\frac{1}{\mu_0}\left(B_a^iB_a^j\right)\,,
    \label{eq:mhdstress}
\end{equation}

The magnitude $\mathcal A_{ab}^i$ is related to the gradient of the interpolator kernel $W_{ab}$ \citep{gsenz_23},  $\eta_a$ is the 2D-Cartesian density (i.e.\ the surface density), $p_a, u_a, a_a^r, a_a^z$ are the gas pressure, internal energy per gram and the radial and vertical accelerations respectively. The parameter $\varepsilon_b$ is one for all particles except for ghost particles located across the symmetry axis for which $\varepsilon_b=-1$. The MHD equations are complemented with the induction equation plus fluid and magnetic dissipative terms in axisymmetric geometry and an efficient divergence-cleaning algorithm \citep{tricco2016, tricco2023}.

\item {\sl Induction equation}

 The induction equation used in the SPH form is \citep{pri12},

\begin{equation}
\frac{d{\bf B}}{dt} = -{\bf B}({\bf\nabla}\cdot{\bf v}) + ({\bf B}\cdot {\bf \nabla}){\bf v}\,,
\label{eq:induction}
\end{equation}
 
\noindent which was adapted to axial geometry in \cite{gsenz_23}, 

\begin{equation}
 \begin{split}
 \frac{d}{dt} 
 &
\left[
\begin{array}{c}
 B^r \\ 
  B^z \\
 B^\varphi 
\end{array}
\right]
= \\
    &\left[
\begin{array}{ccccc}
-\left(\frac{\partial v^z}{\partial z}+\frac{v^r}{r}\right) & \frac{\partial v^r}{\partial z}& -\frac{v^\varphi}{r} \\
\frac{\partial v^z}{\partial r}& -\left(\frac{\partial v^r}{\partial r}+\frac{v^r}{r}\right) & 0\\
\frac{\partial v^\varphi}{\partial r}&\frac{\partial v^\varphi}{\partial z} &-\left(\frac{\partial v^r}{\partial r}+\frac{\partial v^z}{\partial z}\right) \\
\end{array}
\right]
\left[
\begin{array}{c}
 B^r \\ 
  B^z \\
 B^\varphi
\end{array}
\right]\,.
\end{split}
\label{eq:matrixind}
\end{equation}  

Thus, for particle $a$, the induction equation is expressed as a linear equation,

\begin{equation}
\frac{d{B^i_{a}}}{dt} = \sum_{j=1}^3 r^{ij} B^j_{a}
\label{eq:linearinduction}
\end{equation}

\noindent where, according to Eq.~\ref{eq:matrixind}, the coefficients $r^{ij}$ depend only on the velocity of the particle and its derivatives. 

\item {\sl Magnetic dissipation}

The magnetic dissipation term, \( \xi_r \nabla^2 \mathbf{B} \), where $\xi_r$ is the electrical resistivity, leads to the generation of heat due to the resistive dissipation of magnetic energy, similar to the Joule effect in classical electrodynamics. 

In this work, we use an expression for the dissipation, $(du/dt)^{\mathrm{diss}}$ that has been shown to be fully compatible with the conservation of energy in presence of shocks \citep{wissing2020, garciasenz2022}. The magnetic dissipation which contributes to the rate of change of internal energy writes, 
\begin{equation}
     \left(\frac{du}{dt}\right)_a^{\rm diss}=-\frac{\pi r_a}{\mu_0\eta_a}\sum_{b=1}^{n_b}\frac{m_b}{\eta_b}~ \frac{\xi_{r,a}+\xi_{r,b}}{\vert s_{ab}\vert}\mathbf B_{ab}^2\left(\hat s^i_{ab}\mathcal{\tilde A}^i_{ab}\right)\,.
    \label{eq:Bdiss1}
\end{equation}
 
In the simulations below, the adopted value of the resistivity $\xi_r$ is  
\begin{equation}
     \xi_r=\frac{1}{2}\alpha_r~v_{\mathrm{sig},B}\vert s_{ab}\vert\,,
     \label{eq:Bdiss2}
\end{equation}
where $s_{ab}$ is the distance between neighbour particles $a$ and $b$ so that the resistivity value is adjusted to the resolution of the code and $\alpha_r$ is a dimensionless coefficient of the order of unity, \citep{price18}. All models shown in Table \ref{tab:table_kepler} were calculated with the default value $\alpha_r=1$ but other values are explored in Sect.\ref{sec:resistivity}. For the signal velocity, $v_{\mathrm{sig},B}$, we take the Alfvén speed, 
\begin{equation}
   v_{\mathrm{sig},B}= v_{A}=\sqrt{\frac{B^2}{\mu_0\rho}}.
\end{equation}

\item {\sl Equation of state and gravity}

 The equation of state, EOS, includes contributions from the ideal gas of ions and electrons as well as from radiation.  At each integration step, the temperature is obtained from the internal energy by means of a standard implicit Newton-Raphson scheme. Gravity from the NS was implemented as a softened point-like force in the Newtonian approach. A softening distance $r_s = 2.5\times 10^6$ cm was used in all calculations. The disc's self-gravity is subdominant and has been neglected.

\item {\sl Boundary conditions and integration scheme} 

 The integration domain is set up as a thin rectangle, moderately larger than the size of the discs in the r and z directions, whose sides function as open boundaries. For example, in our reference calculation, $x_d=25$, the box sides are at coordinates  $(500, \pm 500)$ km  and $(6200, \pm 500)$ km respectively, but these reduce to  $(25, \pm 50)$ km, $(2100, \pm 50)$ km when $x_d=0.5$. Once a SPH particle crosses the open box boundaries, it is automatically removed from the system. We note that although the computational domain does not extend exactly to $r=0$, firstly because the NS has a finite size but also to avoid numerical problems which could stop the calculations,  the related figures include $r=0$ as a reference. In these figures, the NS is usually indicated by a large dot.

 The MHD equations are integrated with a second-order accurate predictor-corrector scheme. The initial conditions, positions and velocity of the SPH particles are those of the rotating discs after relaxation. As mentioned at the end of Sect.~\ref{sec:B-disk-interaction}, to check the stability of the disc we consider the time-averaged values of Deutsch solution for each component of the magnetic field, $<B_i^2>^{1/2}$, which have an undefined sign. Therefore, this choice of magnetic field is energetically orientated and the calculations work on an energy basis. In the simulations, the averaged pulsar radiation flashes the disc at t=0 and subsequently deactivates. From then on, each component of the magnetised disc evolves according to the axisymmetric SPMHD equations, Eqs. \ref{eq:densitymhd}, \ref{eq:mhdaccel_r}, \ref{eq:mhdaccel_z}, \ref{eq:mhdaccel_phi} and \ref{eq:energy}, together with the induction equation, Eq.~\ref{eq:linearinduction}, 

\begin{equation}
    B^i = B^i (t=0)+ \int_{0}^t \sum_{j=1}^3 r^{ij}B^i~dt\,,
   \label{eq:Int_induc1} 
\end{equation}

\noindent where ${\bf B} (t=0)$ is the average magnetic field of the NS at each particle at the initial instant. The resulting particle motion is driven mainly by heating caused by ohmic dissipation. An alternative procedure for updating ${\bf B}$, is described in Sect.~\ref{sec:induction_b} which, however, produces fairly similar results to those obtained with Eq.~\ref{eq:Int_induc1} during the time elapsed by the simulations. 

\end{itemize}

\section{Disc initial models}
\label{sec:Initial models}

In many binary systems that undergo mass transfer, the expelled material has sufficient angular momentum to form an accretion disc. This is the case for LMXBs and transitional pulsars in the disc state. In these cases, the companion star is usually a low-mass main sequence star and accretion takes place by means of Roche-Lobe overflow. The structure of the disc around the pulsar can be roughly modelled with the analytical thin disc approximation \citep{Shakura}. It should be noted that the stability of an accretion disk depends on the chosen disc scale height, which in turn affects physical processes such as magneto-rotational instability (MRI) or thermal-viscous instability. Thin discs, such as those used in this work ($H/R < 0.05$), are more unstable, less massive, and store less angular momentum than thick discs ($H/R \simeq 0.5$). Thin discs are therefore more fragile and easy to alter by an external perturbation.

\begin{figure}
\centerline{\includegraphics[width=0.52\textwidth]{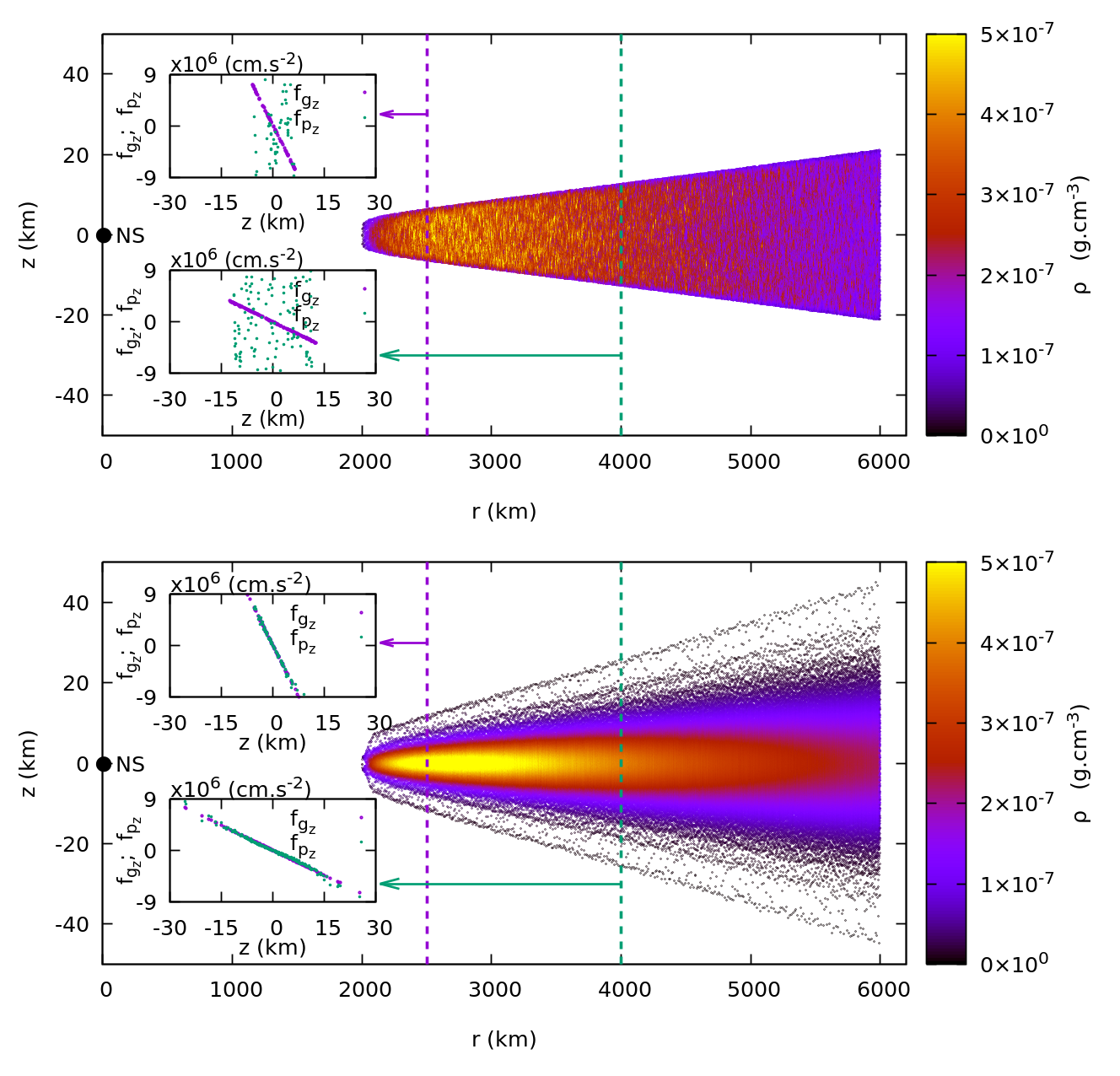}}
     
    \caption{Density distribution of the disc with inner point at $R_{\rm in}= 2000$ km from the pulsar without the magnetic field. {\it Upper panel} shows the disc before relaxation. {\it Lower panel} shows the disc after relaxation. The $z$-component of gravity and gradient of pressure forces along the dotted line at $r=4000$  km and $r=2500$ km are shown in the sub-figures. }
    \label{relaxation}
\end{figure}

To build the initial models used in this work, we first calculate an analytical model of the thin disc following the expressions by \cite{frank02} and with the input parameters summarised in Table~\ref{tab:table1}. The data in Table~\ref{tab:table1} correspond to the well-known tMSP PSR J1023+0038 \citep{archibald2009} which is the prototype of tMSP considered in this work. We produce a suite of 1D disc models at distances $x_{\rm d}$ = $R_{\rm in} / R_{\rm LC}$ ($x_{\rm d}$ = 0.5, 1, 2,  6, 25) from the NS. To have a good resolution, the disc is cut at a distance $R_{\rm out}$ from the NS that is much smaller than the actual size of the disc,   $D_{\rm disc}$, typically $R_{\rm out}/D_{\rm disc} \simeq 2\%$ with the disc size set to its circularization radius \citep{frank02}. In this way, only the innermost part of the disc, closest to the compact object, is resolved by the simulations\footnote{The fraction of the disc considered in this work is, however, considerably larger than that used in other simulations, as for example, \cite{parfrey2017, parfrey2017b}.}. We then map the 1D distributions into a 2D-axisymmetric distribution of SPH particles.
The ensuing discs are further relaxed to remove the excess of spurious numerical noise with the procedure described in Appendix \ref{AppA}. 
A representative example of the relaxation procedure is shown in Fig.~\ref{relaxation} which depicts the density colour-map of the disc before (upper panel) and after (lower panel) the relaxation process is applied to the $R_{\rm in}/R_{\rm LC}=25$ configuration. As can be seen, the disc reacts to maintain vertical equilibrium, resulting in a thicker structure that satisfies the balance between gravitational and pressure forces along the z-axis, $f_{g_z}= -\frac{1}{\rho}\frac{\partial p}{\partial z}$ everywhere. This is shown in the small sub-figures inside the colour-map boxes depicting gravitational $f_{g_z}$ and pressure $f_{p_z}=-\frac{1}{\rho}\frac{\partial p}{\partial z}$ forces along the lines $r=2500, 4000$ km respectively. Unlike the initial disc, which is not in vertical equilibrium, the particle distribution after relaxation is fairly well balanced. 

Finally, the stability of the discs is checked by running the MHD code with the magnetic field switched off during a time $\tau \gg P_{\rm K}$ where $P_{\rm K}$ is the Keplerian period at $R_{\rm in}$, which is denoted $P_{\rm in}$ hereafter. The stability is analysed by monitoring the density and thermal history of several bunches of particles located at different positions along the symmetry axis of the disc, as shown in the two lower rows of sub-figures in Fig.~\ref{multiplot_noB}.
Note that the relative density fluctuations, with respect $\rho(t=0)$, when the pulsar magnetic field is ``off'' are small, within $\simeq 10\%$ level and usually lower. The temperature fluctuates slightly more because the artificial viscosity transforms the numerical noise into internal energy and the oscillation occasionally climbs to $20\%$ in the ``middle'' region of the disc. Also note that the elapsed time covered by the simulations with $\mathbf B=0$ is much larger than that with $\mathbf B\ne 0$ reported below.
The same three regions (close, middle, and far) are used later to track the evolution of the discs once the pulsar radiation is turned on.     

\begin{table}
        \centering
        \caption{Main pulsar parameters used in this work, adapted to the canonical tMSP PSR~J1023+0038. From left to right: name, spin period, light cylinder radius, surface magnetic field strength and magnetic field inclination. We use a NS mass of 1.4~M$_{\sun}$ in all simulations.}
        \begin{tabular}{ccccc} 
                \hline
                MSP & $P$ & $R_{\rm LC}$ & $B_0$ & $\xi$\\
                  & ms & km & G & deg.\\ 
                \hline
                \hline
 PSR J1023+0038& $1.69$ & $80.7$ &$9.6\times 10^7$ & 0$^\circ$, 10$^\circ$, 20$^\circ$, 30$^\circ$\\

                \hline
        \end{tabular}   
        \label{tab:table1}
\end{table}

\section{Results}
\label{sec:calculations}

We performed several simulations of the pulsar-disc interaction where we employed the vacuum solution of Deutsch (1955) which makes a continuous transition from the near zone dipole fields to the far zone radiative fields and modeled the disc with smoothed particle magnetohydrodynamics. We compared our results with the previous analytical work with the final
aim of understanding the tMSP state transitions
 As shown in Fig.~\ref{power_1}, for a given accretion rate,  the outcome of the interaction is determined by the location of the inner radius of the disc, $R_{\rm in}$, and the tilt angle, $\xi$ between the magnetic pole and the spin axis. Consequently, we explore several initial configurations that combine these two parameters and analyse their impact on the stability of the disc by direct MHD simulations with the axisymmetric code. Tables \ref{tab:table1} and \ref{tab:table_kepler} summarise the features of the initial scenarios.

\begin{figure*}
\includegraphics[width=1\textwidth, angle=0.0,origin=c]{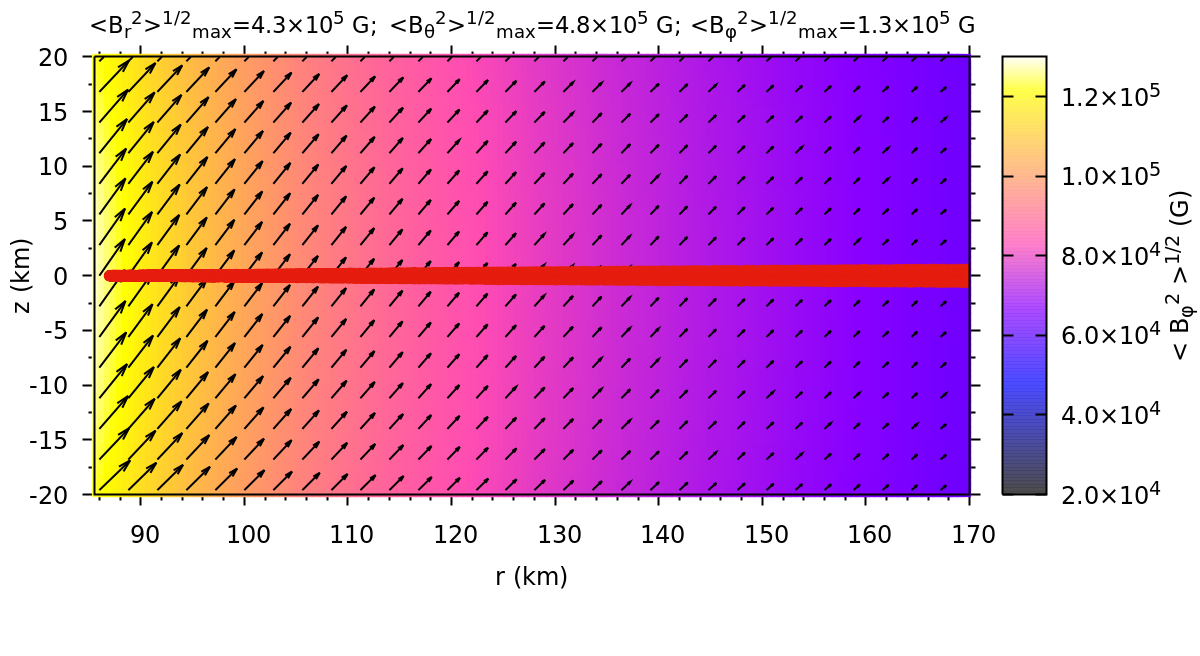}
\includegraphics[width=1\textwidth, angle=0.0,origin=c]{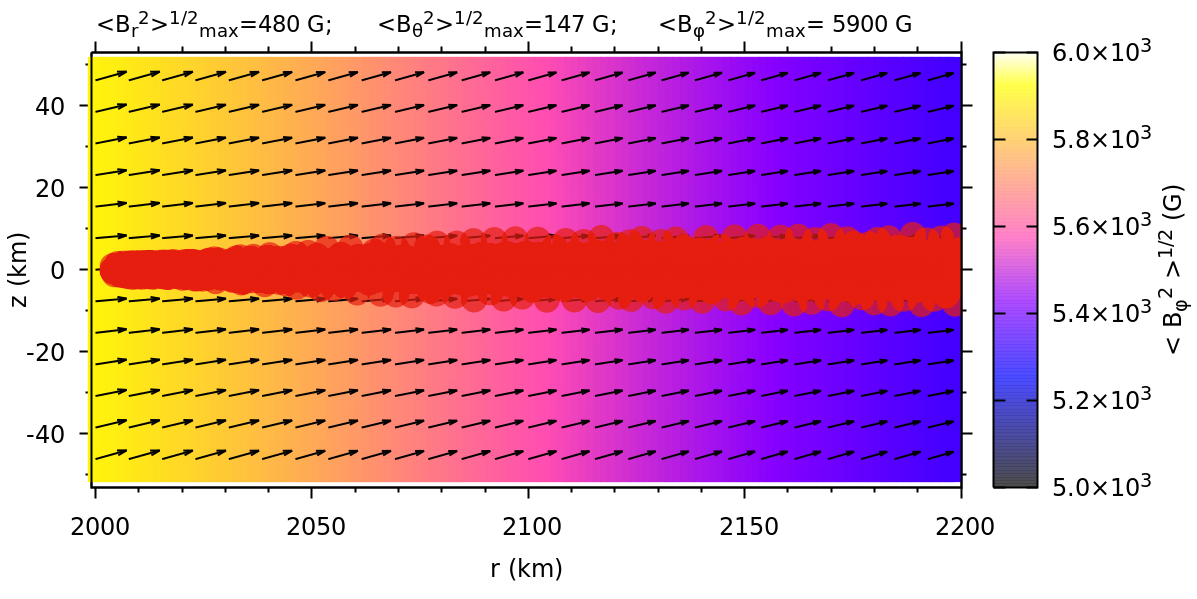}
    \caption{Averaged components of the magnetic field at t=0 and in a small region around $R_{\rm in}$ when $\xi = 20^{\circ}$ and for cases $x_d=1$ (upper panel) and $x_d= 25$ (bottom panel) in Table \ref{tab:table_kepler}. The toroidal field is color-coded and the maximum values of the three components are given as a reference. The silhouette of the discs (in red) is also shown for reference. }
    \label{Bvectors}
\end{figure*}

 Figure \ref{Bvectors} shows a vector map of the time-averaged components of the magnetic field,  $< B_i^2>^{1/2}$, at t=0  for inclined dipoles, $\xi = 20^{\circ}$, and for cases $x_d=1$ (top panel) and $x_d=25$ (bottom panel). For clarity,  only a small region around $R_{\rm in}$ is shown. The black vectors indicate the orientation of the magnetic field in the (r, z) plane, while the azimuthal component is colour-coded. Around the light-cylinder position, $x_d=1$, the polar component slightly dominates the averaged field. Far from the light cylinder, $x_d= 25$, the field practically aligns with the surface of the disc (in red in the figures) and is fully dominated by the azimuthal component of the field. 

 We first describe and discuss the case $x_{\rm d}= R_{\rm in}/R_{\rm LC}=25$ in more detail because the disc appears to be greatly altered after a few rotational periods. Once the full discussion of this reference calculation is complete, we turn to the other cases at lower $x_{\rm d}$ that are described in less detail.

\begin{table}
        \centering
        \caption{Main parameters of the discs used in our simulations, labeled by the value of the inner disc radius in light cylinder units ($x_{\rm d} \equiv R_{\rm in}/R_{\rm LC}$). We assume in all cases a viscosity parameter $\alpha$=0.5 and a mass accretion rate $\dot M$=$10^{13}$~g~s$^{-1}$. The second and third columns show the inner, $R_{\rm in}$ and outer $R_{\rm out}$ disc radius, respectively, fourth and fifth indicate the corresponding Keplerian periods $P_{\rm in}$ and $P_{\rm out}$. The last column shows the number of SPH particles used in each simulation (``25-hr'' is the  high-resolution simulation performed as numerical convergence test, see Sec.~\ref{sec:convergence}).}
        \begin{tabular}{cccccc} 
                \hline
                Model & $R_{\rm in}$  & $R_{\rm out}$  & $P_{\rm in}$ & $P_{\rm out}$ & N-SPH \\ 
                 $x_{\rm d}$ & $R_{\rm LC}$ & $R_{\rm LC}$ & s & s & particles\\
                \hline      
$25$    & 24.8 & 74.3 & 1.3   & 6.7 & $2\times 10^5$ \\
$25$-hr & 24.8 & 74.3 & 1.3   & 6.7 & $8\times 10^5$ \\
$6$      & 5.8  & 49.6 & 0.16  & 3.7 & $4\times 10^5$ \\
$2$ &2.02 & 26.8& 0.032& 1.5& $4\times 10^5$ \\
$1$     & 1.1  & 24.8 & 0.01  & 1.3 & $1.5\times 10^6$  \\
$0.5$   & 0.52 & 24.8 & 0.004 & 1.3 & $1.5\times 10^6$ \\

                \hline
        \end{tabular}   
        \label{tab:table_kepler}
\end{table}

\subsection{Reference calculation: $R_{\rm in} = 2000$ km ($\simeq 25~R_{\rm LC}$)}
\label{sec:reference}

\begin{figure*}
\centering
\includegraphics[scale=0.30]{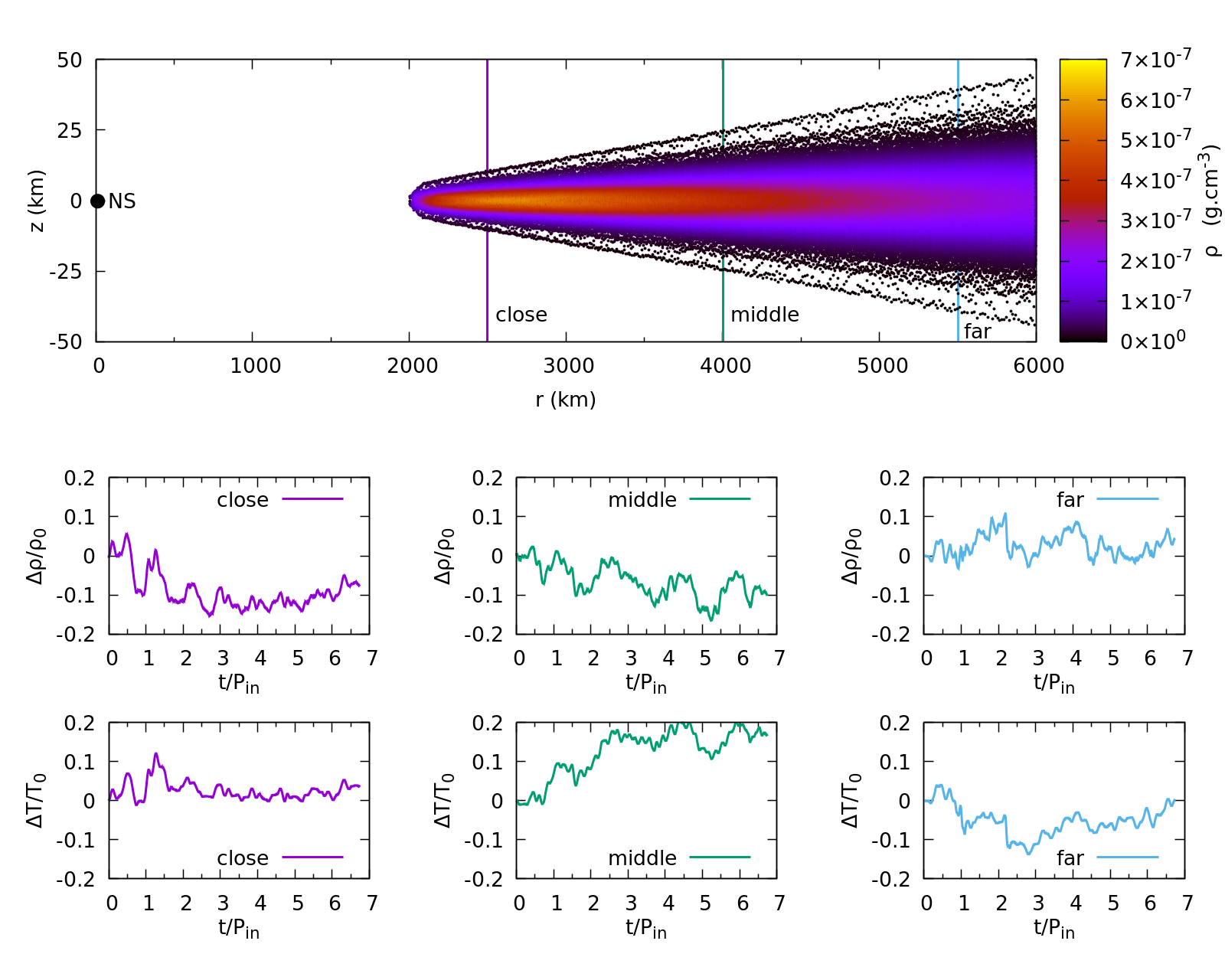}

    \caption{Density distribution (top panel) across the disc at the beginning of the simulation time and for the $x_{\rm d}=25$ model.  Evolution of averaged density and temperature (middle and bottom panels respectively) of the, $z\simeq 0$, ``close'', ``middle'' and ``far'' particle groups when $\mathbf B=0$  as a function of the elapsed time normalised to P$_{\rm in}$. The three vertical lines in the first panel indicate the position of the three groups of particles used as tracers of the stability of the disc (see Sect. \ref{sec:reference})}
    \label{multiplot_noB}
\end{figure*}

This case corresponds to the dimensionless distance $x_{\rm d}=25$ shown in the first row of Table~\ref{tab:table_kepler}. Before starting any of the simulations, it is crucial to check that the disc remains in dynamic equilibrium when $B=0$ for long enough times $\tau = t/P_{\rm in} \gg 1$. For the disc at $x_{\rm d}=25$ this initial state is shown in Figure \ref{multiplot_noB}.  The disc is then suddenly illuminated with the pulsar's radiation and its evolution is calculated for as many orbital periods of its inner point as possible. Practical reasons, however, introduce severe limitations to the total time covered by the simulations because the evaporation of many of the SPH particles at the disc's surface gives birth to an extended diluted cloud of plasma difficult to handle numerically. Eventually, some SPH particles in that cloud become isolated and the simulation stops. This can be mitigated by removing these few ``misbehaving particles'' from the simulation, allowing longer times before the calculation stops. In the $x_{\rm d}=25$ ($R_{\rm in}=2000$ km) model, any particle crossing $ r = 500$ km,  $z = \pm~ 500$ km is removed automatically from the simulation box. 

\begin{figure*}
\centering
\includegraphics[width=1\textwidth]{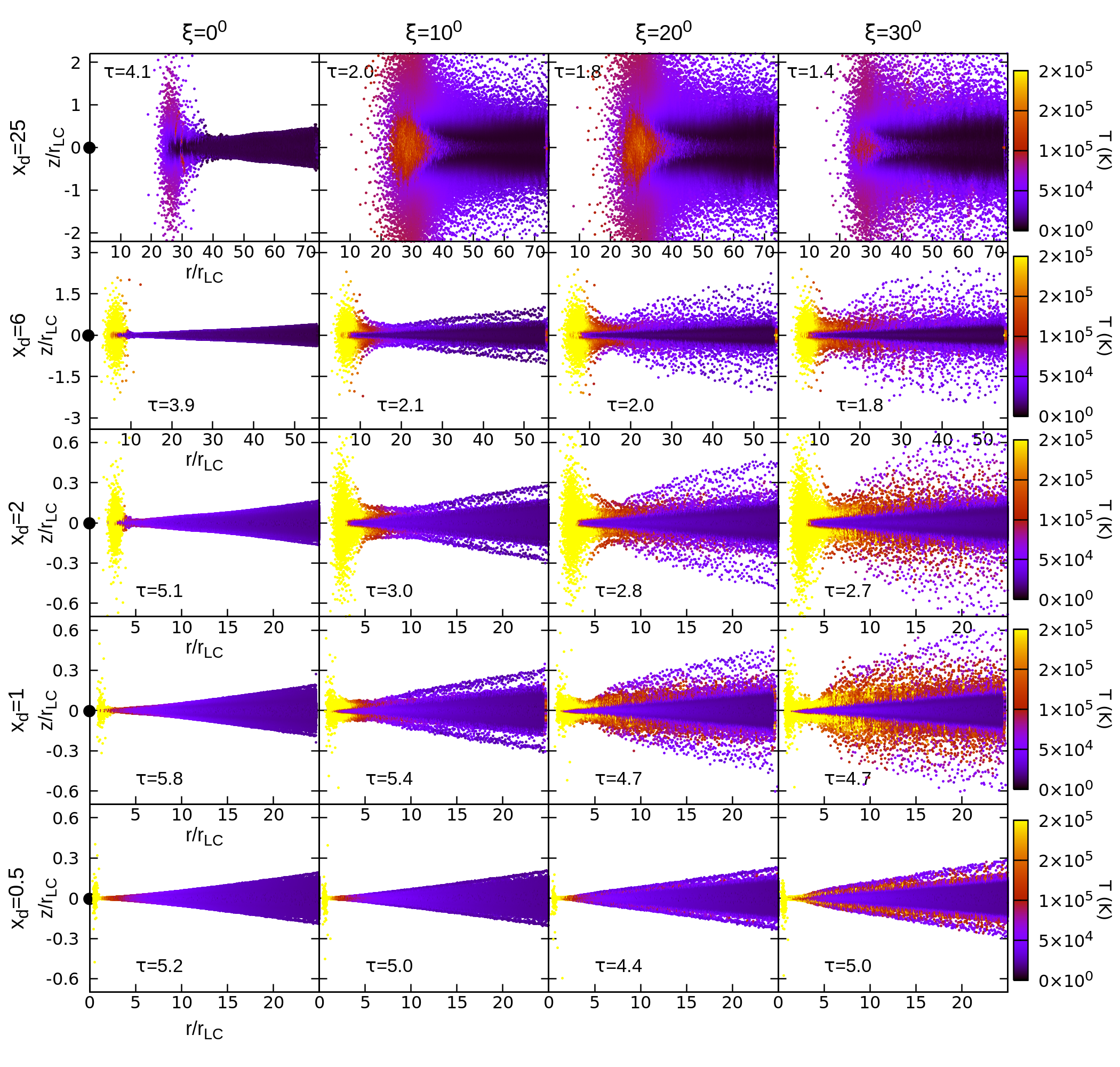}
    \caption{Temperature colour-map of all calculated models around the final time $\tau=t_{\rm f}/P_{\rm in}$ shown in each panel.  Axis coordinates have been normalized to $r_{LC}$. The big dot indicates the position of the NS.}
    \label{Tcolormap_cases}
\end{figure*}

The qualitative behaviour of the disc after switching on the magnetic field is depicted in the first row of Fig.~\ref{Tcolormap_cases}. Each of the four snapshots corresponds to different inclination angles of the magnetic axis of the NS. First of all, it is worth highlighting the very different behaviour between aligned and oblique magnetic rotators. According to Fig.~\ref{deustch_1} the electromagnetic radiation from the aligned, $\xi=0\degree$,  rotator is basically dipolar and affects only a tiny amount of mass around the inner edge of the disc. A small change in the magnetic inclination $\xi$ causes the radial and especially azimuthal components to grow rapidly (see the last row of panels in Fig.~\ref{deustch_1}). These components have a large impact on the part of the disc covered by the simulation, being severely altered in a couple of orbital periods at $\xi=10\degree$. Increasing the value of $\xi$ speeds up the destruction of this fraction of the disc. Therefore, and in agreement with \cite{eksi05} we conclude that the aligned rotator solution for the radiation-disc interaction is unstable when $x_{\rm d}\gg 1$, because a slight change in the magnetic angle induces large alterations of the innermost regions of the disc.

\begin{figure*}
\centering
\includegraphics[width=1\textwidth]{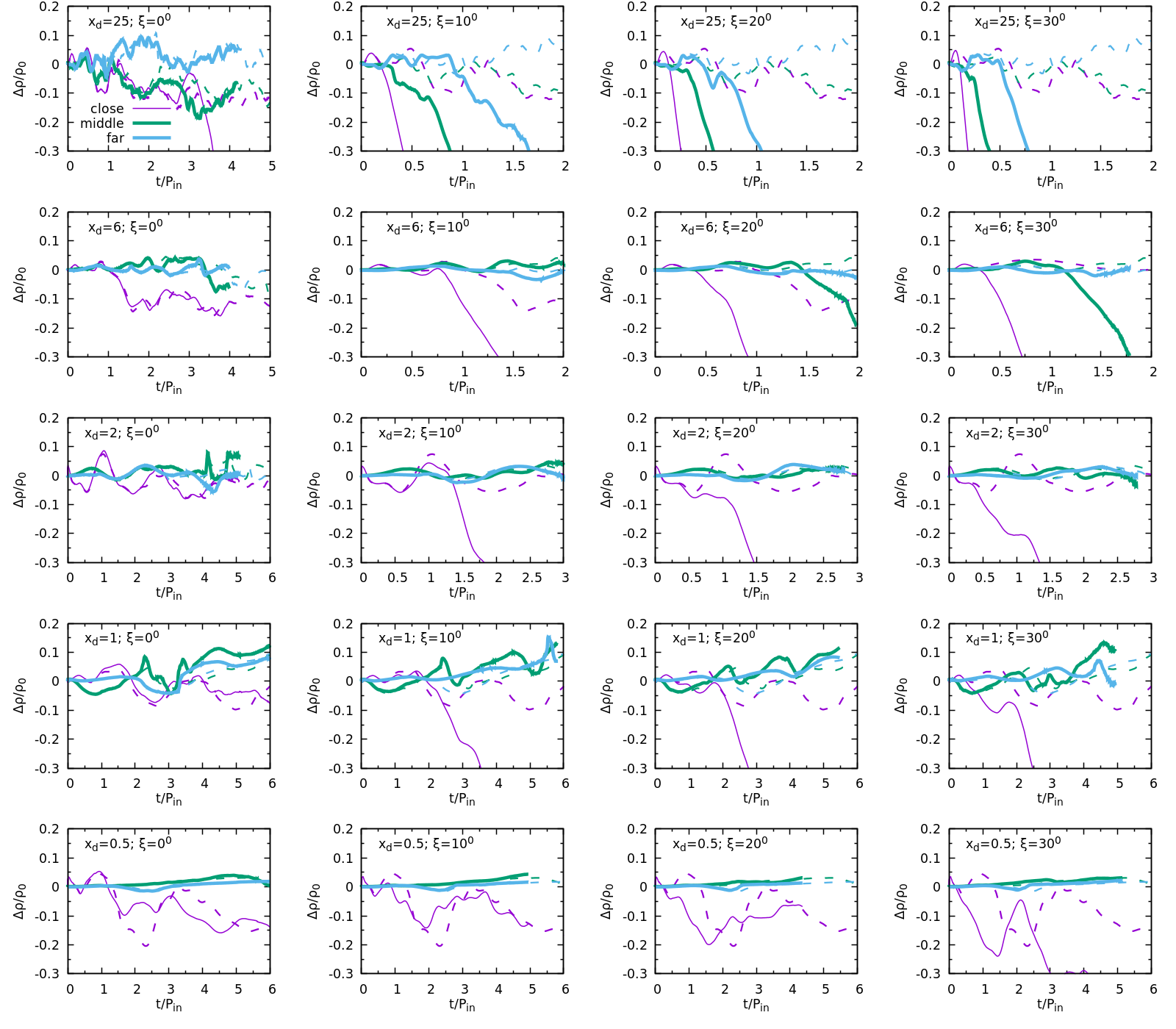}
    \caption{Evolution of $\frac{\Delta\rho}{\rho_0}$ of the calculated models. Continuum lines are for the three groups of particles located along the disc's axis but at different distances from the NS. The ``close'', ``middle'' and ``far'' groups of particles are located at $z=0$ and $r$-coordinates (2500, 4000, 5500) km  ($x_{\rm d}=25$); (1000, 2000, 3500) km ($x_{\rm d}=6$); (350, 1000, 1500) km ($x_{\rm d}=2); $ (200, 800, 1000) km ($x_{\rm d}=1)$ and (100, 800, 1500) km ($x_{\rm d}=0.5)$, respectively. Dashed lines show the evolution of the same groups of particles but with the magnetic field turned off for comparative purposes.} 
    \label{deltarho_rho0}
\end{figure*}

The above conclusion is supported by a quantitative analysis of the behaviour of different groups of disc particles spread at different radial distances from the NS. Figure \ref{deltarho_rho0} shows the normalised deviations of density with respect to its initial values, of the average of three bunches of particles. These are labelled ``close'', ``middle'' and ``far'', respectively, in the figures. The first row in Fig.~\ref{deltarho_rho0}   corresponds to the $x_{\rm d}=25$ case where these groups are initially located at $(2500,0); (4000,0); (5500,0)$ km, from the NS.  On another note, there is a clear correlation between the percent change in $\rho$ with the adopted value of $\xi$ but with the oblique rotators showing considerable variations in the three groups of particles. Especially relevant are the large variations shown by the `middle' and `far' groups of particles in the first row of plots in  Fig.~\ref{deltarho_rho0} which strongly points to the destruction of the simulated part of the disc.    

\begin{figure}
\centering
\includegraphics[width=0.50\textwidth]{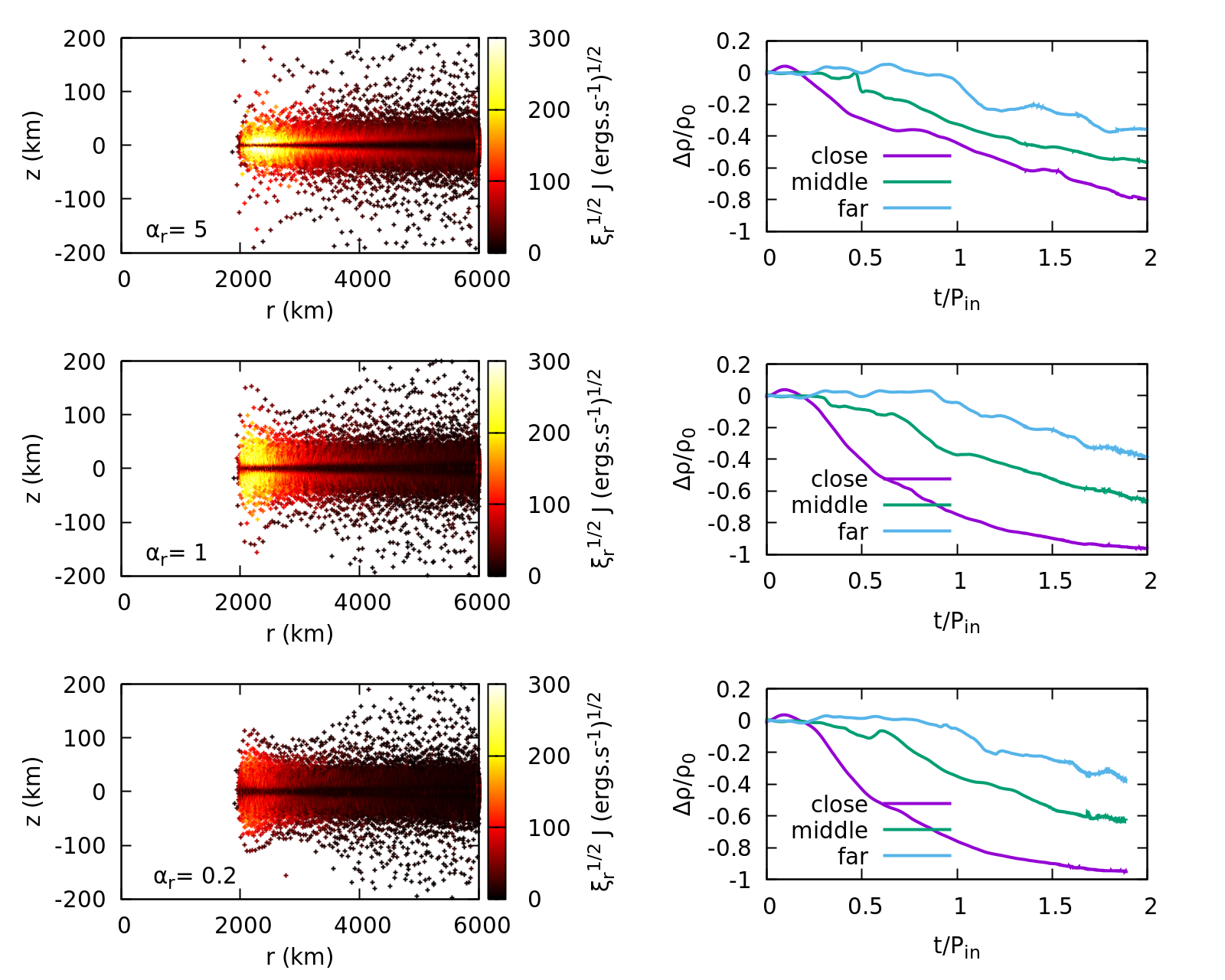}
    \caption{The left column of panels shows the colour-map of the magnitude $\sqrt{\xi_r} \vert \mathbf J\vert$ giving the heat power when squared, for three values of the resistivity parameter, $\alpha_r=0.2, 1, 5$ (from bottom to top) at the normalised time t/P$_{\rm in} = 0.7$. The right column shows the time evolution of $\frac{\Delta\rho}{\rho_0}$ for the three choices of $\alpha_r$ and for the three groups of disc-particles chosen as a tracer.}
    \label{resistivity}
\end{figure}

\subsubsection{Impact of the value of the resistivity parameter}
\label{sec:resistivity}

 The amount of heat dissipation is mainly controlled by the plasma resistivity, $\xi_r$, which operates on microscopic length scales. These scales are often too small to be resolved by current MHD codes. In numerical applications, however, the effect of the physical resistivity is mimicked by a numerical resistivity that adapts the size of the dissipation region to the local resolution of the code. In SPMHD codes, this is done with Eq.~\ref{eq:Bdiss1} which scales the dissipation to the size of the smoothing-length $h$ by conveniently adjusting the parameter $\alpha_r$ in Eq.~\ref{eq:Bdiss2}.  We have changed that parameter over a generous range, $0.2\le \alpha_r\le 5$, to estimate the sensitivity and robustness of the results with the adopted numerical dimensionless parameter $\alpha_r$. To this end, we take the magnetic field  $B_0=9.6\times 10^7$~G, $\xi=10\degree$ and consider three different values of the resistivity parameter, $\alpha_r=0.2, \alpha_r=1$ and $\alpha_r=5$ with $\alpha_r=1$ being the default value used in the simulations. 
 
 Figure \ref{resistivity} summarises the main results of this verification. The first row of panels depict the colour-maps of the variable $\sqrt{\xi_r} \vert\mathbf J\vert$ where $\xi_r$ is the local resistivity given by Eq.~\ref{eq:Bdiss2} and $\mathbf J = \mathbf\nabla\times \mathbf B/\mu_0$, so its square gives the heat dissipation power. Overall, the impact of increasing the resistivity is to produce more heat but less expansion of the disc's shells which is also evident from the behaviour of the three groups of tracer particles (defined in Sect.~\ref{sec:reference}) shown in the second column of panels in Fig.~\ref{resistivity}. A high value of $\xi_r$ increases the diffusion of $B$, causing a larger spread of the region where magnetic energy is transformed into heat which can occur, for example, in the presence of shocks and stabilises the magnetised plasma. Nevertheless, the evolution of the disc when $\alpha_r=0.2$ does not differ appreciably from the default case $\alpha_r=1$, which is encouraging. However, a much larger resistivity value, such as that obtained with $\alpha_r > 5-10$, may have an appreciable impact on the evolution of the disc.      

\subsubsection{Numerical Convergence}
\label{sec:convergence}

\begin{figure}
\centering
\includegraphics[width=0.48\textwidth]{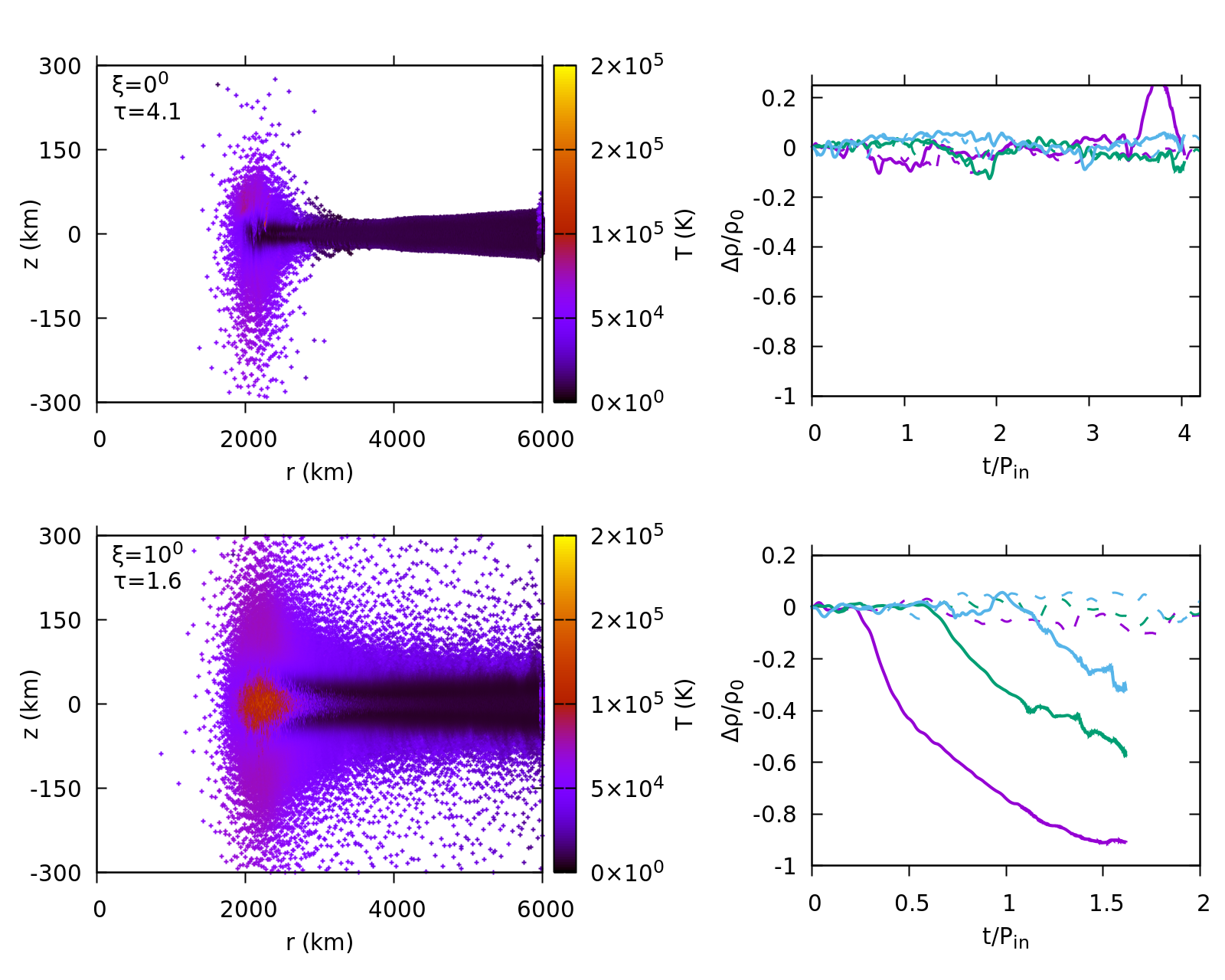}
    \caption{Model $x_{\rm d}=25$ calculated with a factor two improved resolution. The top row of panels depicts the evolution of the $\xi=0^0$ aligned rotator while  the second row is for the angle $\xi=10^0$ at times $\tau=t/P_{\rm in}$. The first column of panels shows the density colour- map and the second tracks the evolution of density in the same three groups of SPH particles as in the first row of Fig.\ref{Tcolormap_cases}. Dashed lines are the trajectories followed by the same groups of particles but calculated with $B=0$~ for comparative purposes.}
    \label{resolution}
\end{figure}

The sensitivity of the simulations to the adopted resolution has been explored simply by increasing the number of SPH particles. We have repeated the calculation with $x_{\rm d}=25$  for the models with $\xi=0^0, 10^0$ but this time using $N=8\times 10^5$ particles (second row in Table \ref{tab:table_kepler}) instead of our default value $N=2\times 10^5$, so that the size of the smoothing length $h$ is halved. Comparing the first two panels of Figs~\ref{Tcolormap_cases} and \ref{deltarho_rho0} with Fig. \ref{resolution}, we notice that they are qualitatively and quantitatively similar, which is reassuring,  but with two minor differences that are worth mentioning. The first is the slight asymmetry in the distribution of the stripped material around the inner edge of the disc shown by the aligned rotator. This is due to the feedback between the numerical noise, which is higher in this region due to surface effects\footnote{It has to be taken into account that each particle configuration was relaxed prior to their  acceptance  as a suitable initial model. However, some residual numerical noise always remains. On another note, the error in estimating gradients and derivatives around the inner edge of the disc is larger because the SPH interpolations are sensitive to particle sampling, which is not smooth in this abrupt frontier region separating matter and vacuum. }, and the stronger magnetic field felt by the plasma. The second difference is that the $\Delta\rho/\rho_0$ profile of the first group of particles (magenta line in the upper right panel) reaches a larger positive peak at $\tau\simeq 4$ than in the low-resolution calculation (first panel in Fig.~\ref{deltarho_rho0}) before starting the downward decompression path. This is most likely due to the focusing effect induced by the reaction of evaporated particles around the disc vertex toward the $r$-axis, which is larger in the high-resolution calculation. 

Overall, the high-resolution calculation led to somewhat lower plasma temperatures. This can be understood in terms of the plasma resistivity, Eq.~\ref{eq:Bdiss2} which becomes less than the default value with the $N=2\times 10^5$ particles used in Sect. \ref{sec:reference} and is in agreement with the results obtained in the previous section,  where the impact of changing the resistivity parameter was addressed.    

\label{convergence}

  \subsection{Cases at different NS-disc distances}

In this Section,  we present the effects of changing the position of the inner radius of the disc. We carry out simulations for $R_{\rm in}$ at the co-rotation radius ($0.5 R_{\rm LC}$), at the light cylinder ($R_{\rm LC}$), and moderately outside the light cylinder radius ($2 R_{\rm LC}, 6 R_{\rm LC}$). The main features of these scenarios are summarised in Table~\ref{tab:table_kepler}. It should be noted that a larger number of SPH particles have been used in these calculations and that the portion of the simulated disc is smaller than in the reference case, especially when $r\le R_{\rm LC}$. This is because the disc becomes sharper as it approaches the LC radius, so more resolution is needed to define it.  It is worth noting that close to the right-most edge of the disc an anomalous temperature feature appears, which is caused by the open boundary conditions used (see, for instance, the last three panels in the second row in Fig.~\ref{Tcolormap_cases}). These features arise because of the back-reaction of the SPH particles that move to the right. However, the numerical artifact affects a negligible amount of mass and does not have an appreciable impact on the outcome of the simulations \footnote{A possible improvement in future implementations of the code is to consider reflective conditions at the right-most edge of the disc. However, these conditions are not straightforward to implement in a SPH code with axial geometry due to the ring-like nature of the SPH particles. }

A summary of the results is shown in Figures~\ref{Tcolormap_cases} and \ref{deltarho_rho0} that include all the calculated models. Moving along each column from top to bottom in these figures provides information on the impact of change $R_{\rm in}$ at constant magnetic inclination $\xi$. Moving along each row from left to right explores the sensitivity of the simulated portion of the disc at constant $R_{\rm in}$ (or $x_{\rm d}$) but different $\xi$.   

The geometry of the disc for aligned rotators, $\xi=0\degree$, is not greatly affected by pulsar radiation, and the disc survives. Only a small amount of the mass of the disc around $R_{\rm in}$ shows signs of instability, but this is a peculiar region
where the numerical errors in the SPH interpolations are expected to be larger.  In general, the SPH  technique has second-order accuracy, and we expect the numerical errors and noise to be relatively small between the regions of the discs. The exception is around the disc's edges and, especially, around $R_{\rm in}$, where particle sampling is less uniform. The numerical noise is greatly reduced during the relaxation of the disc and only the region around $R_{\rm in}$ ultimately remains slightly unstable. This  unstable region typically extends over a region several  times the value of the smoothing length $h$,  so it involves too small an amount of mass to have a significant impact on the results. Moreover, the most problematic simulations at $x_d=0.5$ and $x_d=1$, where the magnetic field strength is higher, were run using many more particles (see Table~\ref{tab:table_kepler}) to minimise the size of such an unstable region around $R_{in}$. 

The most obvious pattern in Fig.~\ref{Tcolormap_cases} is the destabilising effect of increasing $\xi$ for constant $x_d$, where higher values of $\xi$ produce more ablation on the surface of the disc. Nevertheless, the behaviour of the disc with increasing $\xi$ is also a sensitive function of distance $x_{\rm d}$. For $x_{\rm d}=25$,
the disc is severely altered in any $\xi > 0\degree$, as suggested by the behaviour of the three tracer regions (continuum lines in Figs.~ \ref{deltarho_rho0}), and was also discussed in Sect.~\ref{sec:reference}. At lower distances ($x_{\rm d}=6$) the disc becomes much more stable but with clear signs of instability for $\xi \gtrapprox 20^0$, as suggested by the temperature colour-maps and the two last panels of Fig.~\ref{deltarho_rho0}.  There, both the `close' and the `middle' tracer regions (continuum magenta and green lines) depart from the stability line of reference determined with $B=0$ (dashed lines). 

As the distance to the NS shortens, $x_{\rm d}=2$, and approaches 
the light cylinder radius $x_{\rm d}\simeq 1$, the disc becomes progressively more stable owing to the prevalence of the polar component of the magnetic field. Still, the radial and azimuthal components of $\mathbf B$ become sufficiently high at $\xi \gtrapprox 20\degree$ (see the first and second rows in Fig.~\ref{deustch_1}) to cause the evaporation of some particles on the surface of the disc, although the disc is altered only on a much smaller scale than in the $x_{\rm d}=6, 25$ calculations.  Unlike the case $x_{\rm d}=6$, the density tracer in the middle region of the disc remained unaffected during the simulation time, suggesting better stability. For discs truncated at the corotation radius at $x_{\rm d}\simeq 0.5$, the axisymmetric simulations suggest a quite stable behaviour, even at the highest inclination angle $\xi=30^\degree$ considered in this work. The last row of panels in Figs.~\ref{Tcolormap_cases} and \ref{deltarho_rho0} supports the above statement. Only a negligible amount of the mass of the disc around $R_{\rm in}$ is ejected up and down and in the NS direction. Hence, our direct numerical simulations indicate that the X-ray state of MSP is only plausible at a sufficiently low radius $ 0.5 R_{\rm LC}\lessapprox r \lessapprox 1 R_{\rm LC}$ . Larger radii are not excluded as long as the magnetic axis is $\xi \lessapprox 10\degree-20\degree$, with the precise value depending on the particular value of $x_{\rm d}$.   

\subsection{Changing the scheme to update the magnetic field}

\label{sec:induction_b}

\begin{figure}
\centering
\includegraphics[width=0.50\textwidth]{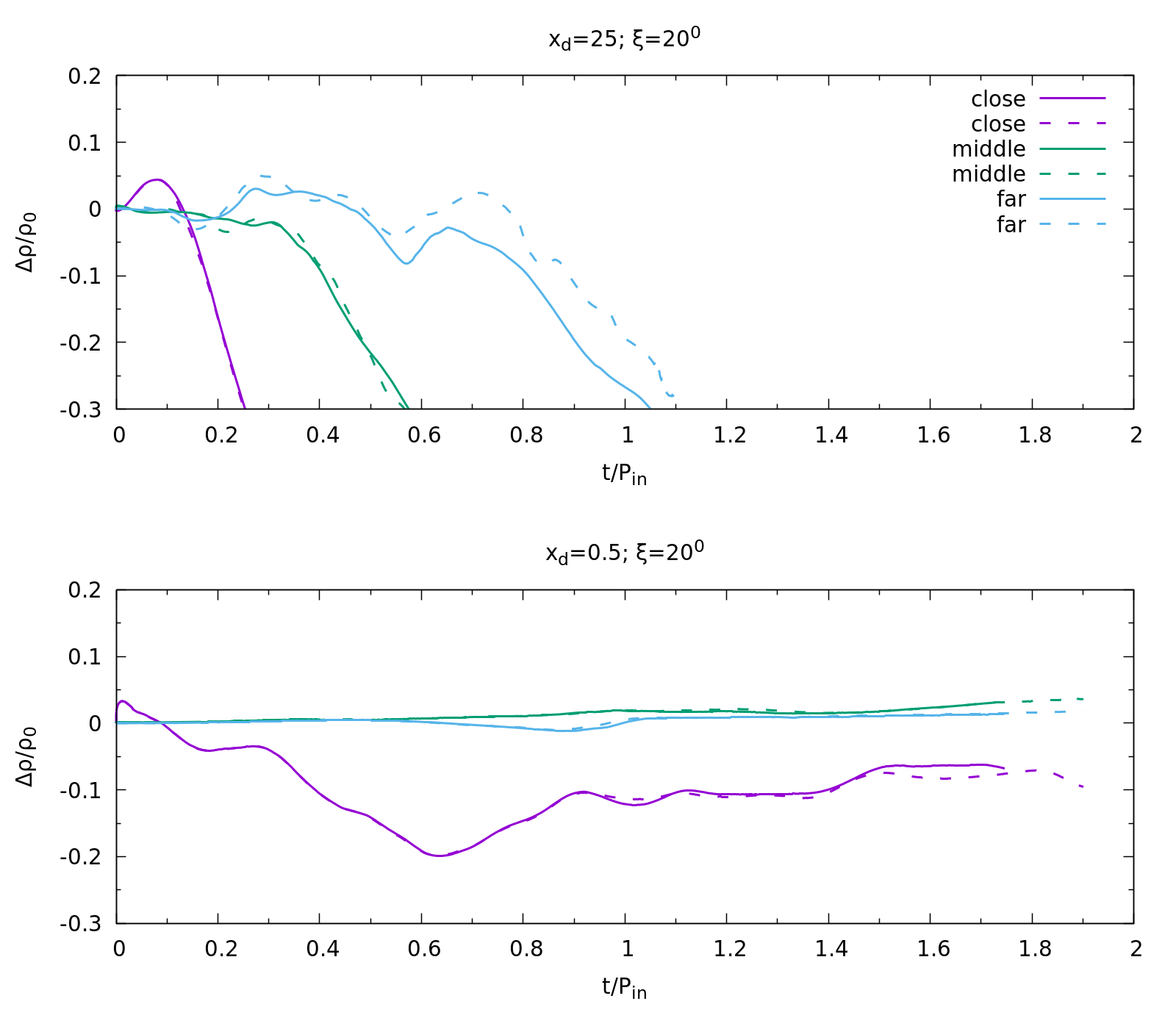}
    \caption{Evolution of $\Delta\rho/\rho_0$ for the three groups of tracer particles calculated with the ${\bf B}$-update schemes given by Eq.~\ref{eq:Int_induc1} (continuum lines) and Eq.~\ref{eq:Int_inducb} (dashed lines).}
    \label{InducTwoSchemes}
\end{figure}

Another procedure for evolving the magnetic field consists of separating the contributions of the external field of the NS, ${\bf B}_{\rm ext}$, from the contribution of magnetic induction ${\bf B}_{\rm ind}$. The evolution of the $i-$component of the magnetic field attached to particles, $B^i$, therefore is, 
    
\begin{equation}
    B^i (t) = B^i_{ext} (t)+ B^i_{ind} (t)
   \label{eq:Int_inducb}\,, 
\end{equation}

\noindent where $B^i_{ind} (t) = B^i_{ind}(t=0)+\int_0^t \sum_{j=1}^3 r^{ij} B^i dt$~ (Eq. \ref{eq:linearinduction}),  with $B_{ind}^i(t=0)=0$  and with $B^i_{ext}(t=0)$ being the average pulsar magnetic field at t=0. This second scheme has the advantage that the magnetic field of the NS is continuously updated along with the displacement of the particles. However, it significantly increases the computational burden because the averages $< B^2 >^{1/2}$ have to be calculated at any integration step and for any particle. Using Eq.~\ref{eq:Int_inducb} we have recalculated some models in Table \ref{tab:table_kepler} and concluded that both schemes lead to similar results. This can be seen in Fig.~\ref{InducTwoSchemes}, which shows the two approaches for the extreme cases $x_d=25$ and $x_d=0.5$ with $\xi=20^{\circ}$. The evolution is fairly similar in the two cases, almost
identical when $x_d$ is low.  

\section{Discussion}
\label{sec:discussion}

Traditionally, the classification of accreting NSs into pulsar (ejector), propeller, and accretor states has been done using energy density (or pressure) arguments relating the electromagnetic pressure to the gas pressure of the disc \citep{1992Lipunov, papitto2022}, which leads to the results shown in Fig.~\ref{power_1} and discussed in Section~\ref{sec:B-disk-interaction}. This procedure was extended by \cite{eksi05} who considered the full solution of the multipolar electromagnetic field radiated by an oblique magnetic rotator obtained by \cite{deutsch55} allowing the regime $R_{\rm in} \gtrsim R_{\rm LC}$ to be investigated. Still, the classification is based on the identity $p_{\rm rad}=p_{\rm gas}$ and does not take into account other details such as the role played by the particular disc structure. This work aims to address this shortcoming and present MHD numerical simulations of the interaction between the electromagnetic radiation from the NS and the inner regions of realistic accretion discs. Because the focus of the interaction is still on the energy balance, we can compare the results of the simulations with those described in the analytical work by \cite{eksi05}.   

To do this, we have placed our results in the ($\xi-x_{\rm d}$) stability diagram obtained by \cite{eksi05}, seeking a comparison between the two methods. As outlined in Sec.~\ref{sec:calculations}, we classify the numerical simulations into two groups, stable and unstable, adopting a simple criterion: we consider the simulated disc to be unstable when the ``middle'' or the ``far'' regions, highlighted with thick continuum lines in Fig.~\ref{deltarho_rho0}, show clear signs of decay with time and with respect to the reference stability line (dashed lines).
This decay can be quantified as a relative change in density of at least 20\%.
Otherwise, the simulation is labelled as stable. This criterion does not take into account the set of SPH particles in the ``close'' region because these are at a low depth and too close to the inner edge of the disc, where the numerical errors are larger and the amount of mass involved is low. The evolution of the ``close'' region is, however, shown by the thin solid magenta line in Fig.~\ref{deltarho_rho0}. As we shall see, the results of such density-based criterion are consistent with the trend observed in the temperature colour-maps (Fig.~\ref{Tcolormap_cases}). 

\begin{figure}
\centering
\includegraphics[width=0.5\textwidth]{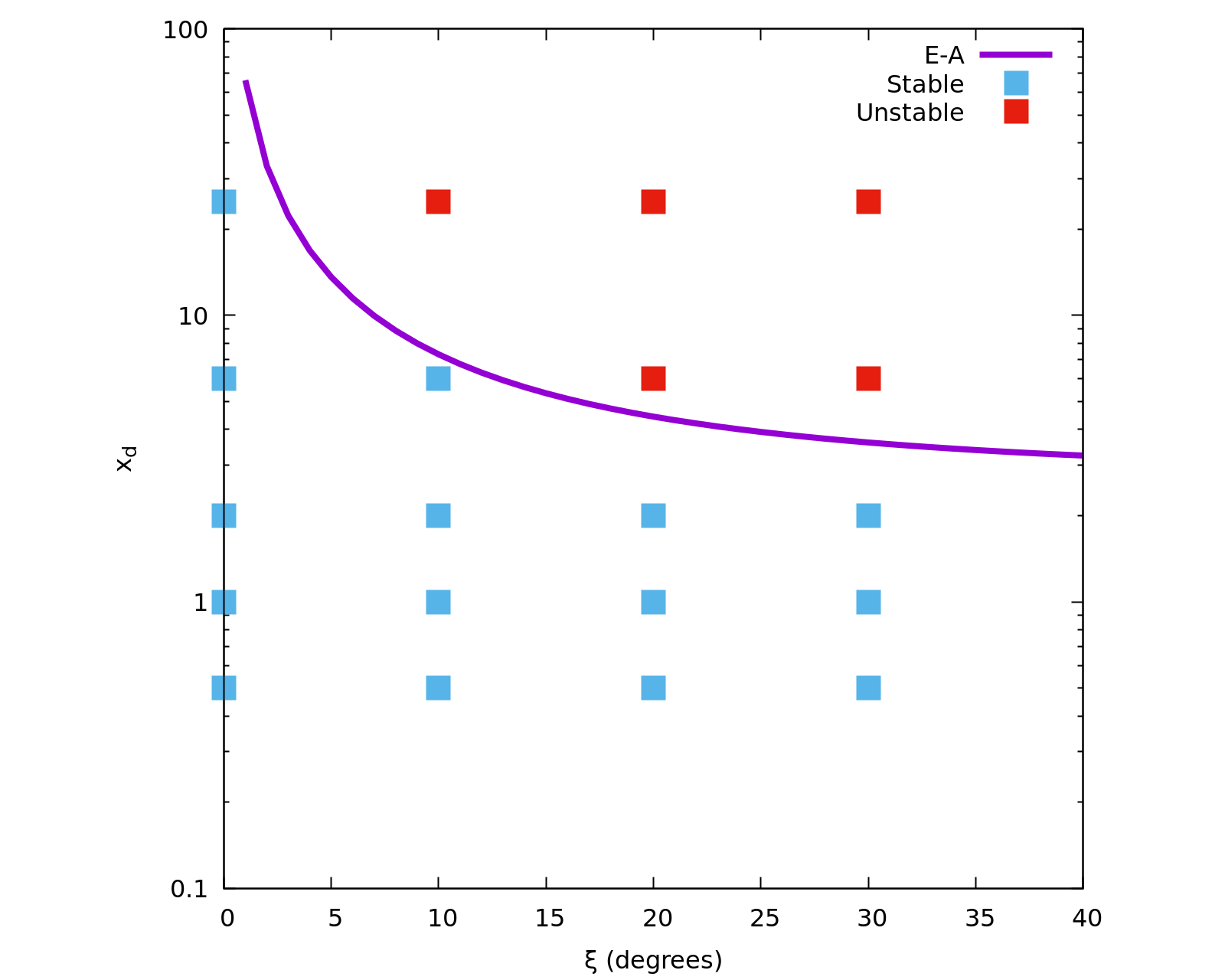}
    \caption{Disc stability in the $\xi-x_d$ diagram. The thick line in magenta is the theoretical line by \citet{eksi05} separating stable (below the line) from unstable (above the line) discs. The squares represent the results from direct numerical simulations in this work. Blue squares indicate a not too much altered, and therefore, stable disc and red squares depict the unstable cases.}
    \label{stability}
\end{figure}

The disc stability regions are shown in Fig.~\ref{stability}, where the solid thick line is the theoretical line by \cite{eksi05} and the coloured squares show our simulations.  As we can see, and despite the very different methods used to address the problem, there is good agreement between the two, with the stable/unstable cases (blue/red squares) lying below/above the stability line, respectively.

For stable cases, mainly with $x_{\rm d}$ reaching a few $R_{\rm LC}$, our simple energy model only provides a general check on disc stability, and the detailed dynamical interaction between the pulsar wind and the accretion flow must be studied with other simulations \citep{parfrey2017,parfrey2017b}.
The disruption of the innermost part of the disc in the unstable cases can be interpreted as the first step towards a transition from the disc to the pulsar state.
In the following, we discuss the implications for transitional MSPs, LMXBs and fallback discs.

Since in a realistic setting a neutron star is surrounded by a magnetized plasma, the vacuum solution we employed has its own limitations in terms of discussing the disc stability. The presence of the corona could influence the stability of the disc in different ways: For example, the MRI \citep{bal91} is sustained in the presence of moderate magnetic fields \citep{bal98}. If the corona imposes a strong large-scale poloidal magnetic field, it can suppress MRI turbulence, leading to a reduction in effective viscosity and mass accretion rate. Since the corona provides an additional source of pressure, it can modify the vertical structure of the disk \citep[see e.g.][]{mis02,jia+19}. The magnetic buoyancy effects can lead to the rise of field lines into the corona which can cause episodic reconnection events, possibly leading to disk instabilities or flaring behaviour \citep{aly90} This also affects the transport of angular momentum if magnetic fields couple the disk and the corona. A strong coronal field can drive magnetised winds, extracting angular momentum from the disk \citep{kad+18} This can affect the stability of the inner disk regions, particularly if a magnetically driven wind depletes material faster than the disk can replenish it. Obviously, studying some of these effects would require three-dimensional simulations which are beyond the scope of the present paper.

\subsection{Transitional Millisecond Pulsars}

We currently know 31 confirmed redback MSPs in the pulsar state, and
only two of them have been observed in the disc state (the so-called
tMSPs).
In most models of the disc state of tMSPs, the inner disc radius is
close to the light cylinder $R_{\rm in} \simeq R_{\rm LC}$ or, say, $x_{\rm d}$ = [0.5-2] \citep{2014Linares, campana2016,veledina2019, linares2022}.
This should correspond to a stable disc configuration, since the disc
state persists for years-decades (Sect.~\ref{sec:intro}).
The angle of inclination of the spin-magnetic field $\xi$ cannot change on such
short timescales\footnote{Timescales for aligning the spin and magnetic axis have been determined by \cite{philippov2014}, for both vacuum and plasma filled isolated pulsars. In the two cases and for $B_0=10^8$~G and $P=1.69$ ms, typical alignment times are enormous, exceeding $10^{10}$ years. However, that value is merely indicative because the estimate did not consider the impact of the surrounding disc.}. Therefore,  changes in magnetic obliquity (horizontal displacements in the $\xi - x_{\rm d}$ diagram, Fig.~\ref{stability}) are not a plausible mechanism to drive the tMSP state transitions.

The inner disc, on the other hand, is much more dynamic \citep[as
  witnessed e.g.\ by MHD simulations of the innermost accretion
  flow][]{parfrey2017} and $R_{\rm in}$ can change on much shorter
timescales, from years to less than a second, facilitating short vertical excursions in Fig.~\ref{stability}.  
Rapid fluctuations of $R_{\rm in}$ were proposed to trigger the fast
($\sim$second) X-ray mode switches, bringing the inner disc
inside/outside the light cylinder \citep{2014Linares}.
Here we propose that long-term fluctuations (months to years) in $R_{\rm in}$ offer a natural way to drive and trigger state transitions in tMSPs, moving between stable and unstable states of the disc.
If $R_{\rm in}$ enters the unstable region, the innermost region of the disc can be fully evaporated or ejected on relatively short timescales, and a transition from the disc to the pulsar state may follow. Our simulations cannot reproduce such $R_{\rm in}$ variability, but next we
discuss the fluctuations required for different magnetic inclinations.

If the magnetic obliquity is small, $\xi \lesssim 5^\circ$, large
fluctuations are needed to move $R_{\rm in}$ to the unstable region.
We estimate that $R_{\rm in}$ should change by a factor $\gtrsim 10$ to
trigger a state transition for such a low $\xi$ (from $x_{\rm d} \simeq$~2 to
20, Figure~\ref{stability}).
If the magnetic obliquity is large, $\xi \gtrsim 20^\circ$, small
fluctuations are sufficient to move $R_{\rm in}$ to the unstable region.
For example, a factor $\simeq$3-4 change in $R_{\rm in}$ would be
sufficient to move the disc to the unstable region and trigger a state
transition at $\xi \simeq 20^\circ$ (roughly from $x_{\rm d} \simeq$~2 to 6, Figure~\ref{stability}).
It is worth noting in this context that the X-ray luminosity of the
disc state is remarkably constant on timescales of years ($L_{\rm X}$
changes by a factor $\lesssim$2, besides X-ray mode switching and
flares; see, e.g., Figure 1 in \citealt{shahbaz2019}).
This suggests that the changes in $R_{\rm in}$ are subtle, in line with
the high $\xi$ scenario for tMSPs.
Thus, we suggest that tMSPs have sizeable obliquities, which allow them to transition between disc and pulsar states with relatively small
fluctuations in $R_{\rm in}$ and $L_{\rm X}$.

\begin{figure}
\centering
\includegraphics[width=0.50\textwidth]{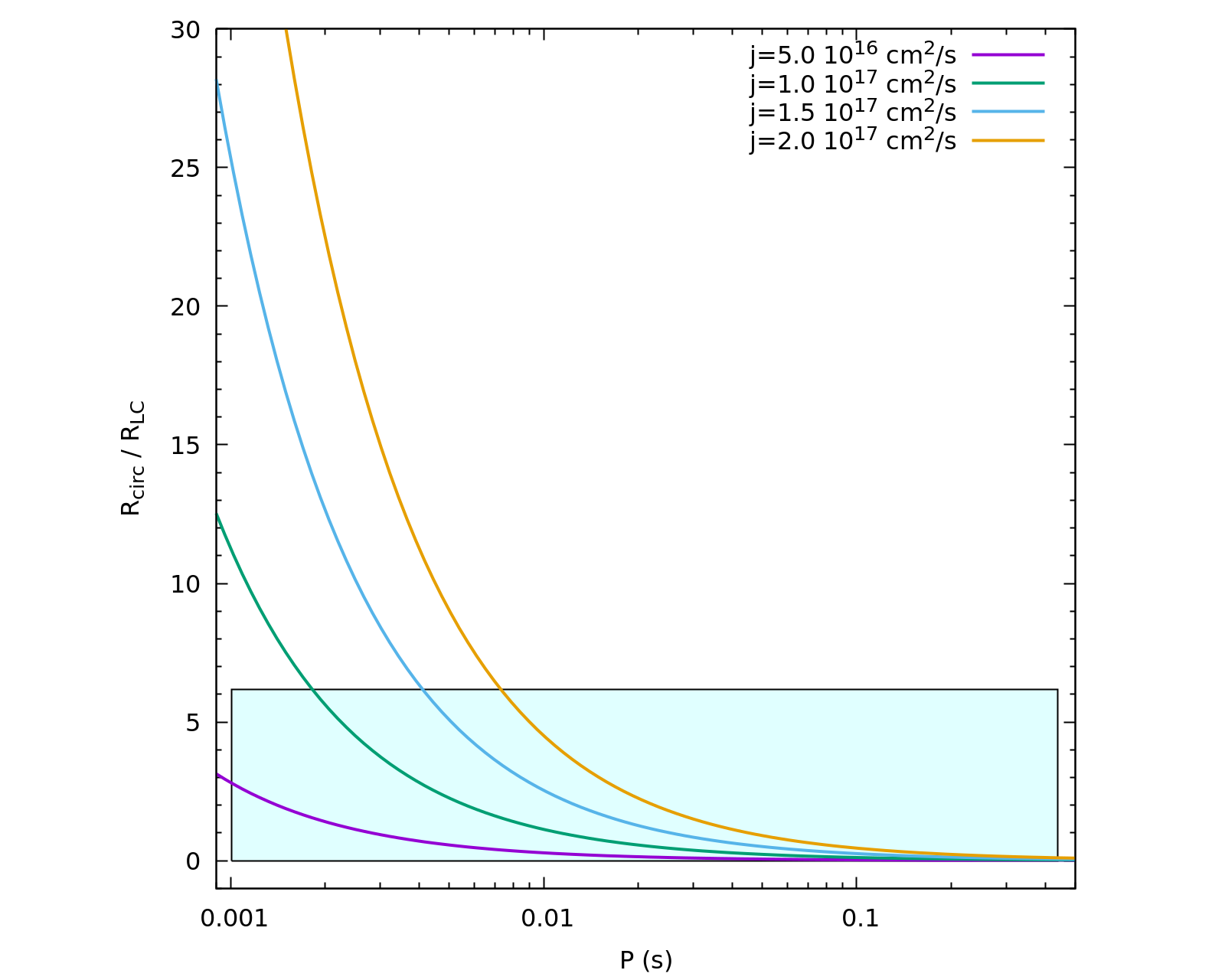}
    \caption{Circularization radius (normalized to $R_{\rm LC}$) of matter in the supernova fallback disc scenario as a function of the rotation period of the neutron star. The circularization radius is given for four choices of the specific angular momentum $j$ of the falling matter. The box in light blue shows the region with the highest probability of disc survival, greater near the bottom of that region.}
    \label{fallback}
\end{figure}
radius
\subsection{Low-mass X-ray binaries}

In this context, it is interesting to compare the long-term stability of $L_{\rm X}$ for tMSPs in the disc state with their evolutionary predecessors: NSs in LMXBs.
Between outburst and quiescence, some NS-LMXBs show variable X-ray luminosity on timescales of weeks-years, in the range
$L_{\rm X} \sim 10^{34}-10^{36}$~erg~s$^{-1}$ \citep{Wijnands13,Bernardini13,Allen15}.
Their X-ray spectrum typically softens with decreasing $L_{\rm X}$ in this range (with the photon index going from $\simeq$1.5 to $\simeq$3; \citealt{2014bLinares,Wijnands15}).
This reveals a fundamental difference in their low-level accretion properties: at low mass accretion rates, accretion discs in tMSPs can persist for years at relatively constant $L_{\rm X}$, while accretion discs in NS-LMXBs are more strongly variable.

Such wider $L_{\rm X}$ changes in LMXBs suggest that their disc inner radius ($x_{\rm d}$) can change widely while the disc remains stable.
Looking back at Figure~\ref{stability}, we see that this can occur at low magnetic obliquity ($\xi$).
Thus, we suggest that NS-LMXBs (at least those showing strong X-ray variability in quiescence) have nearly aligned magnetic and spin axes, $\xi \lesssim 10^\circ$.
%


\subsection{Supernova fallback discs}

The regime investigated here ($R_{\rm in} \simeq R_{\rm LC}$) is also relevant for the putative supernova fallback discs around young pulsars \citep{mic81,mic88}. The presence of such discs was proposed as an explanation for the braking indices \citep{men+01,cal+13}, the distribution of pulsars in the $P-\dot{P}$-diagram \citep{alp+01}, the diversity of young neutron star families \citep{cha+00, alp01}, and the existence of pulsar jets \citep{bla04}. The first debris disc around a young neutron star, a magnetar, was discovered by \citet{wan+06} who modelled the infrared spectrum with a passive disc beyond the light cylinder radius. The infrared and optical spectra were interpreted with a viscous active disc by \citet{ert+07}. The presence of a passive disc around the Vela pulsar beyond the radius of the light cylinder was proposed by \citet{dan+11} based on the excess infrared emission. A fallback disc model in which the full spectrum could be explained by an active disc protruding the light cylinder was proposed by \citet{ozs+14}. Since the Vela pulsar is relatively close, and more distant debris discs are not easily detected due to the dust abundant in the galactic disc midplane, the presence of a disc around the Vela pulsar would imply that these discs are ubiquitous. Most recently, fallback discs were proposed \citep{ron+22, gen+23} as an agent to slow down ultra-long period pulsars \citep{hur+22}.
Recent work by \citet{jan+22} and \citet{bar+22} suggests the significant role of fallback material with angular momentum around young neutron stars for addressing the spin-kick alignment and the initial spin-up of magnetars to the periods required by the dynamo action. 

Unlike the steady, long-lived discs considered in this work, those fallback discs are transient events. Still, we can connect our results to that scenario by assuming $R_{\rm in}\simeq R_{\rm circ}$, where $R_{\rm circ} = j^2/GM_{\rm NS}$ is the circularization radius  and $j$ is the specific angular momentum of the gas in the fallback disc scenario. Figure \ref{fallback} shows the position of the circularization radius normalised to the light-cylinder radius as a function of the rotation period of the NS and for different values of $j$. The coloured box indicates the region where, according to our results, the disc has a chance of surviving, more likely near the bottom of it, close to $x_{\rm d}\simeq R_{\rm circ}/ R_{\rm LC}=1$. Numerical simulations favour $10^{16}\lesssim j\lesssim 10^{17}$ cm$^2$~s$^{-1}$ \citep{jan+22} for which a large fraction of the curves lie in the stability region. Furthermore, our results suggest that between $1\lesssim x_{\rm d}\lesssim 6$ these discs are more likely to survive if the tilt angle $\xi$ is small or zero.

\section{Conclusions}
\label{sec:conclusions}

The interaction between the electromagnetic radiation emitted by a NS and the nearby accretion disc has been simulated and analysed in light of an energetic approach that considers the time-constant squared average of the electromagnetic field. Our novel SPMHD axisymmetric simulations suggest that the inner regions of the disc are hardly disturbed when the inner radius of the disc, $R_{\rm in}$, is similar to or lower than the radius of the light cylinder. At larger distances, the stability depends on the interplay between the position of $R_{\rm in}$ and the value of the magnetic inclination angle. At sufficiently large values of the inner radius, $x_{\rm d} \gtrsim 10$,  the disc is always unstable, which facilitates its truncation. 
The only exception is when the magnetic axis is exactly aligned to the spin axis. Our results also suggest that changes in the state of the pulsar could be induced by short displacements of the inner radius of the disc,  between $x_{\rm d} \simeq 2 - 7$, provided that the inclination of the magnetic axis is $\xi \gtrsim 20^\circ$. 

Despite the different approaches, our results agree well with those of \cite{eksi05} and highlight both the importance of including the full electromagnetic solution of Deutsch \citep{deutsch55} in numerical simulations of recycled MSP and the correct description of plasma resistivity.  These issues are especially relevant for better understanding the physical basis behind the tMSP phenomena. This work also shows that the SPMHD technique can be safely used to study the MSP scenario,  but much work remains to be done. The SPH method has previously been used to study the hydrodynamic interaction between the wind produced by the central object and the disc \citep[e.g.,][]{okazaki2011} although without the presence of the magnetic field. A natural extension is to drop the energy approach used in this work and introduce the magnetised pulsar wind into the scheme to simulate the full MHD interaction with the disc in three dimensions. Another issue worth exploring is the impact of pulsar radiation in previously magnetised discs, although obtaining stable, relaxed disc structures under such conditions could be difficult.

\section*{Acknowledgements}

The authors thank the anonymous referee for insightful comments and suggestions that helped to improve this manuscript. E.V. acknowledges support from the Spanish Ministry
of Economy and Competitiveness (MINECO) under grant AYA2017-
86274-P and the Spanish Ministry of Science and Innovation/State Research Agency (MICINN/AEI) under grant PID2020-117252GB-I00.   
D.G-S. acknowledges the support of the Spanish MINECO grant PID2020-117252GB-100 and the AGAUR/Generalitat de Catalunya grant SGR-386/2021. M.L. thanks M. Armas-Padilla and D. Misra for discussions on low-level accretion and magnetic obliquity evolution, respectively. M.L. also acknowledges funding from the European Research Council (ERC) under the European Union’s Horizon 2020 research and innovation programme (grant agreement No. 101002352). KYE thanks the Physics Department of Sultan Qaboos University for their hospitality.

\section*{Data Availability}

The data supporting this article will be shared on reasonable request to the corresponding author. The code used to carry out the simulations, Axis-SPHYNX is publicly available at:~\href{https://github.com/realnewton/AxisSPHYNX}{Axis-SPHYNX download}



i

\begin{thebibliography}{99}

\makeatletter
\relax
\def\mn@urlcharsother{\let\do\@makeother \do\$\do\&\do\#\do\^\do\_\do\%\do\~}
\def\mn@doi{\begingroup\mn@urlcharsother \@ifnextchar [ {\mn@doi@} {\mn@doi@[]}}
\def\mn@doi@[#1]#2{\def\@tempa{#1}\ifx\@tempa\@empty \href {http://dx.doi.org/#2} {doi:#2}\else \href {http://dx.doi.org/#2} {#1}\fi \endgroup}
\def\mn@eprint#1#2{\mn@eprint@#1:#2::\@nil}
\def\mn@eprint@arXiv#1{\href {http://arxiv.org/abs/#1} {{\tt arXiv:#1}}}
\def\mn@eprint@dblp#1{\href {http://dblp.uni-trier.de/rec/bibtex/#1.xml} {dblp:#1}}
\def\mn@eprint@#1:#2:#3:#4\@nil{\def\@tempa {#1}\def\@tempb {#2}\def\@tempc {#3}\ifx \@tempc \@empty \let \@tempc \@tempb \let \@tempb \@tempa \fi \ifx \@tempb \@empty \def\@tempb {arXiv}\fi \@ifundefined {mn@eprint@\@tempb}{\@tempb:\@tempc}{\expandafter \expandafter \csname mn@eprint@\@tempb\endcsname \expandafter{\@tempc}}}

\bibitem[\protect\citeauthoryear{{Abolmasov}, {Biryukov}  \& {Popov}}{{Abolmasov} et~al.}{2024}]{abo+24}
{Abolmasov} P.,  {Biryukov} A.,   {Popov} S.~B.,  2024, \mn@doi [Galaxies] {10.3390/galaxies12010007}, \href {https://ui.adsabs.harvard.edu/abs/2024Galax..12....7A} {12, 7}

\bibitem[\protect\citeauthoryear{{Allen}, {Linares}, {Homan}  \& {Chakrabarty}}{{Allen} et~al.}{2015}]{Allen15}
{Allen} J.~L.,  {Linares} M.,  {Homan} J.,   {Chakrabarty} D.,  2015, \mn@doi [\apj] {10.1088/0004-637X/801/1/10}, \href {https://ui.adsabs.harvard.edu/abs/2015ApJ...801...10A} {801, 10}

\bibitem[\protect\citeauthoryear{{Alpar}}{{Alpar}}{2001}]{alp01}
{Alpar} M.~A.,  2001, \mn@doi [\apj] {10.1086/321393}, \href {http://adsabs.harvard.edu/abs/2001ApJ...554.1245A} {554, 1245}

\bibitem[\protect\citeauthoryear{{Alpar}, {Cheng}, {Ruderman}  \& {Shaham}}{{Alpar} et~al.}{1982}]{alpar1982}
{Alpar} M.~A.,  {Cheng} A.~F.,  {Ruderman} M.~A.,   {Shaham} J.,  1982, \mn@doi [\nat] {10.1038/300728a0}, \href {https://ui.adsabs.harvard.edu/abs/1982Natur.300..728A} {300, 728}

\bibitem[\protect\citeauthoryear{{Alpar}, {Ankay}  \& {Yazgan}}{{Alpar} et~al.}{2001}]{alp+01}
{Alpar} M.~A.,  {Ankay} A.,   {Yazgan} E.,  2001, \mn@doi [\apjl] {10.1086/323140}, \href {http://adsabs.harvard.edu/abs/2001ApJ...557L..61A} {557, L61}

\bibitem[\protect\citeauthoryear{{Aly} \& {Kuijpers}}{{Aly} \& {Kuijpers}}{1990}]{aly90}
{Aly} J.~J.,  {Kuijpers} J.,  1990, \aap, \href {https://ui.adsabs.harvard.edu/abs/1990A&A...227..473A} {227, 473}

\bibitem[\protect\citeauthoryear{{Archibald} et~al.,}{{Archibald} et~al.}{2009}]{archibald2009}
{Archibald} A.~M.,  et~al., 2009, \mn@doi [Science] {10.1126/science.1172740}, \href {https://ui.adsabs.harvard.edu/abs/2009Sci...324.1411A} {324, 1411}

\bibitem[\protect\citeauthoryear{{Balbus} \& {Hawley}}{{Balbus} \& {Hawley}}{1991}]{bal91}
{Balbus} S.~A.,  {Hawley} J.~F.,  1991, \mn@doi [\apj] {10.1086/170270}, \href {https://ui.adsabs.harvard.edu/abs/1991ApJ...376..214B} {376, 214}

\bibitem[\protect\citeauthoryear{{Balbus} \& {Hawley}}{{Balbus} \& {Hawley}}{1998}]{bal98}
{Balbus} S.~A.,  {Hawley} J.~F.,  1998, \mn@doi [Reviews of Modern Physics] {10.1103/RevModPhys.70.1}, \href {https://ui.adsabs.harvard.edu/abs/1998RvMP...70....1B} {70, 1}

\bibitem[\protect\citeauthoryear{{Barr{\`e}re}, {Guilet}, {Reboul-Salze}, {Raynaud}  \& {Janka}}{{Barr{\`e}re} et~al.}{2022}]{bar+22}
{Barr{\`e}re} P.,  {Guilet} J.,  {Reboul-Salze} A.,  {Raynaud} R.,   {Janka} H.~T.,  2022, \mn@doi [\aap] {10.1051/0004-6361/202244172}, \href {https://ui.adsabs.harvard.edu/abs/2022A&A...668A..79B} {668, A79}

\bibitem[\protect\citeauthoryear{{Bernardini}, {Cackett}, {Brown}, {D'Angelo}, {Degenaar}, {Miller}, {Reynolds}  \& {Wijnands}}{{Bernardini} et~al.}{2013}]{Bernardini13}
{Bernardini} F.,  {Cackett} E.~M.,  {Brown} E.~F.,  {D'Angelo} C.,  {Degenaar} N.,  {Miller} J.~M.,  {Reynolds} M.,   {Wijnands} R.,  2013, \mn@doi [\mnras] {10.1093/mnras/stt1741}, \href {https://ui.adsabs.harvard.edu/abs/2013MNRAS.436.2465B} {436, 2465}

\bibitem[\protect\citeauthoryear{{Blackman} \& {Perna}}{{Blackman} \& {Perna}}{2004}]{bla04}
{Blackman} E.~G.,  {Perna} R.,  2004, \mn@doi [\apjl] {10.1086/381802}, \href {http://adsabs.harvard.edu/abs/2004ApJ...601L..71B} {601, L71}

\bibitem[\protect\citeauthoryear{{Campana}, {Colpi}, {Mereghetti}, {Stella}  \& {Tavani}}{{Campana} et~al.}{1998}]{campana1998}
{Campana} S.,  {Colpi} M.,  {Mereghetti} S.,  {Stella} L.,   {Tavani} M.,  1998, \mn@doi [\aapr] {10.1007/s001590050012}, \href {https://ui.adsabs.harvard.edu/abs/1998A&ARv...8..279C} {8, 279}

\bibitem[\protect\citeauthoryear{{Campana}, {Coti Zelati}, {Papitto}, {Rea}, {Torres}, {Baglio}  \& {D'Avanzo}}{{Campana} et~al.}{2016}]{campana2016}
{Campana} S.,  {Coti Zelati} F.,  {Papitto} A.,  {Rea} N.,  {Torres} D.~F.,  {Baglio} M.~C.,   {D'Avanzo} P.,  2016, \mn@doi [\aap] {10.1051/0004-6361/201629035}, \href {https://ui.adsabs.harvard.edu/abs/2016A&A...594A..31C} {594, A31}

\bibitem[\protect\citeauthoryear{{Cavelan}, {Cabez{\'o}n}, {Grabarczyk}  \& {Ciorba}}{{Cavelan} et~al.}{2020}]{Cavelan2020}
{Cavelan} A.,  {Cabez{\'o}n} R.~M.,  {Grabarczyk} M.,   {Ciorba} F.~M.,  2020, in PASC '20: Proceedings of the Platform for Advanced Scientific Computing ConferenceJune 2020. p.~11 (\mn@eprint {arXiv} {2005.02656}), \mn@doi{10.1145/3394277.3401855}

\bibitem[\protect\citeauthoryear{{Cerutti}, {Philippov}  \& {Spitkovsky}}{{Cerutti} et~al.}{2016}]{cerutti2016}
{Cerutti} B.,  {Philippov} A.~A.,   {Spitkovsky} A.,  2016, \mn@doi [\mnras] {10.1093/mnras/stw124}, \href {https://ui.adsabs.harvard.edu/abs/2016MNRAS.457.2401C} {457, 2401}

\bibitem[\protect\citeauthoryear{{Cerutti}, {Philippov}  \& {Dubus}}{{Cerutti} et~al.}{2020}]{cerutti2020}
{Cerutti} B.,  {Philippov} A.~A.,   {Dubus} G.,  2020, \mn@doi [\aap] {10.1051/0004-6361/202038618}, \href {https://ui.adsabs.harvard.edu/abs/2020A&A...642A.204C} {642, A204}

\bibitem[\protect\citeauthoryear{{Chatterjee}, {Hernquist}  \& {Narayan}}{{Chatterjee} et~al.}{2000}]{cha+00}
{Chatterjee} P.,  {Hernquist} L.,   {Narayan} R.,  2000, \mn@doi [\apj] {10.1086/308748}, \href {http://adsabs.harvard.edu/abs/2000ApJ...534..373C} {534, 373}

\bibitem[\protect\citeauthoryear{{Danilenko}, {Zyuzin}, {Shibanov}  \& {Zharikov}}{{Danilenko} et~al.}{2011}]{dan+11}
{Danilenko} A.~A.,  {Zyuzin} D.~A.,  {Shibanov} Y.~A.,   {Zharikov} S.~V.,  2011, \mn@doi [\mnras] {10.1111/j.1365-2966.2011.18753.x}, \href {https://ui.adsabs.harvard.edu/abs/2011MNRAS.415..867D} {415, 867}

\bibitem[\protect\citeauthoryear{{Das} \& {Porth}}{{Das} \& {Porth}}{2024}]{das24}
{Das} P.,  {Porth} O.,  2024, \mn@doi [\apjl] {10.3847/2041-8213/ad151f}, \href {https://ui.adsabs.harvard.edu/abs/2024ApJ...960L..12D} {960, L12}

\bibitem[\protect\citeauthoryear{{Das}, {Porth}  \& {Watts}}{{Das} et~al.}{2022}]{das+22}
{Das} P.,  {Porth} O.,   {Watts} A.~L.,  2022, \mn@doi [\mnras] {10.1093/mnras/stac1817}, \href {https://ui.adsabs.harvard.edu/abs/2022MNRAS.515.3144D} {515, 3144}

\bibitem[\protect\citeauthoryear{{Davidson} \& {Ostriker}}{{Davidson} \& {Ostriker}}{1973}]{davidson1973}
{Davidson} K.,  {Ostriker} J.~P.,  1973, \mn@doi [\apj] {10.1086/151897}, \href {https://ui.adsabs.harvard.edu/abs/1973ApJ...179..585D} {179, 585}

\bibitem[\protect\citeauthoryear{{Deutsch}}{{Deutsch}}{1955}]{deutsch55}
{Deutsch} A.~J.,  1955, Annales d'Astrophysique, \href {https://ui.adsabs.harvard.edu/abs/1955AnAp...18....1D} {18, 1}

\bibitem[\protect\citeauthoryear{{Ek{\c{s}}i} \& {Alpar}}{{Ek{\c{s}}i} \& {Alpar}}{2005}]{eksi05}
{Ek{\c{s}}i} K.~Y.,  {Alpar} M.~A.,  2005, \mn@doi [\apj] {10.1086/425959}, \href {https://ui.adsabs.harvard.edu/abs/2005ApJ...620..390E} {620, 390}

\bibitem[\protect\citeauthoryear{{Ertan}, {Erkut}, {Ek{\c s}i}  \& {Alpar}}{{Ertan} et~al.}{2007}]{ert+07}
{Ertan} {\"U}.,  {Erkut} M.~H.,  {Ek{\c s}i} K.~Y.,   {Alpar} M.~A.,  2007, \mn@doi [\apj] {10.1086/510303}, \href {http://adsabs.harvard.edu/abs/2007ApJ...657..441E} {657, 441}

\bibitem[\protect\citeauthoryear{{Frank}, {King}  \& {Raine}}{{Frank} et~al.}{2002}]{frank02}
{Frank} J.,  {King} A.,   {Raine} D.~J.,  2002, {Accretion Power in Astrophysics: Third Edition}.
{Cambridge University Press}

\bibitem[\protect\citeauthoryear{{Garc{\'\i}a-Senz}, {Cabez{\'o}n}, {Blanco-Iglesias}  \& {Lor{\'e}n-Aguilar}}{{Garc{\'\i}a-Senz} et~al.}{2020}]{2020Garcia}
{Garc{\'\i}a-Senz} D.,  {Cabez{\'o}n} R.~M.,  {Blanco-Iglesias} J.~M.,   {Lor{\'e}n-Aguilar} P.,  2020, \mn@doi [\aap] {10.1051/0004-6361/201936837}, \href {https://ui.adsabs.harvard.edu/abs/2020A&A...637A..61G} {637, A61}

\bibitem[\protect\citeauthoryear{{Garc{\'\i}a-Senz}, {Wissing}  \& {Cabez{\'o}n}}{{Garc{\'\i}a-Senz} et~al.}{2022}]{garciasenz2022}
{Garc{\'\i}a-Senz} D.,  {Wissing} R.,   {Cabez{\'o}n} R.~M.,  2022, \mn@doi [arXiv e-prints] {10.48550/arXiv.2206.05324}, \href {https://ui.adsabs.harvard.edu/abs/2022arXiv220605324G} {p. arXiv:2206.05324}

\bibitem[\protect\citeauthoryear{{Garc{\'\i}a-Senz}, {Wissing}, {Cabez{\'o}n}, {Vurgun}  \& {Linares}}{{Garc{\'\i}a-Senz} et~al.}{2023}]{gsenz_23}
{Garc{\'\i}a-Senz} D.,  {Wissing} R.,  {Cabez{\'o}n} R.~M.,  {Vurgun} E.,   {Linares} M.,  2023, \mn@doi [\mnras] {10.1093/mnras/stac3328}, \href {https://ui.adsabs.harvard.edu/abs/2023MNRAS.518.4115G} {518, 4115}

\bibitem[\protect\citeauthoryear{{Gen{\c{c}}ali}, {Ertan}  \& {Alpar}}{{Gen{\c{c}}ali} et~al.}{2023}]{gen+23}
{Gen{\c{c}}ali} A.~A.,  {Ertan} {\"U}.,   {Alpar} M.~A.,  2023, \mn@doi [\mnras] {10.1093/mnrasl/slac164}, \href {https://ui.adsabs.harvard.edu/abs/2023MNRAS.520L..11G} {520, L11}

\bibitem[\protect\citeauthoryear{{Ghosh} \& {Lamb}}{{Ghosh} \& {Lamb}}{1979a}]{gho79a}
{Ghosh} P.,  {Lamb} F.~K.,  1979a, \mn@doi [\apj] {10.1086/157285}, \href {http://adsabs.harvard.edu/abs/1979ApJ...232..259G} {232, 259}

\bibitem[\protect\citeauthoryear{{Ghosh} \& {Lamb}}{{Ghosh} \& {Lamb}}{1979b}]{gho79b}
{Ghosh} P.,  {Lamb} F.~K.,  1979b, \mn@doi [\apj] {10.1086/157498}, \href {http://adsabs.harvard.edu/abs/1979ApJ...234..296G} {234, 296}

\bibitem[\protect\citeauthoryear{{Goldreich} \& {Julian}}{{Goldreich} \& {Julian}}{1969}]{gol69}
{Goldreich} P.,  {Julian} W.~H.,  1969, \mn@doi [\apj] {10.1086/150119}, \href {https://ui.adsabs.harvard.edu/abs/1969ApJ...157..869G} {157, 869}

\bibitem[\protect\citeauthoryear{{Guerra}, {Meliani}  \& {Voisin}}{{Guerra} et~al.}{2024}]{guerra2024}
{Guerra} C.,  {Meliani} Z.,   {Voisin} G.,  2024, \mn@doi [arXiv e-prints] {10.48550/arXiv.2407.14842}, \href {https://ui.adsabs.harvard.edu/abs/2024arXiv240714842G} {p. arXiv:2407.14842}

\bibitem[\protect\citeauthoryear{{Hayashi}, {Shibata}  \& {Matsumoto}}{{Hayashi} et~al.}{1996}]{Hay+96}
{Hayashi} M.~R.,  {Shibata} K.,   {Matsumoto} R.,  1996, \mn@doi [\apjl] {10.1086/310222}, \href {https://ui.adsabs.harvard.edu/abs/1996ApJ...468L..37H} {468, L37}

\bibitem[\protect\citeauthoryear{{Hurley-Walker} et~al.,}{{Hurley-Walker} et~al.}{2022}]{hur+22}
{Hurley-Walker} N.,  et~al., 2022, \mn@doi [\nat] {10.1038/s41586-021-04272-x}, \href {https://ui.adsabs.harvard.edu/abs/2022Natur.601..526H} {601, 526}

\bibitem[\protect\citeauthoryear{{Illarionov} \& {Sunyaev}}{{Illarionov} \& {Sunyaev}}{1975}]{1975Illarionov}
{Illarionov} A.~F.,  {Sunyaev} R.~A.,  1975, \aap, \href {https://ui.adsabs.harvard.edu/abs/1975A&A....39..185I} {39, 185}

\bibitem[\protect\citeauthoryear{{Janka}, {Wongwathanarat}  \& {Kramer}}{{Janka} et~al.}{2022}]{jan+22}
{Janka} H.-T.,  {Wongwathanarat} A.,   {Kramer} M.,  2022, \mn@doi [\apj] {10.3847/1538-4357/ac403c}, \href {https://ui.adsabs.harvard.edu/abs/2022ApJ...926....9J} {926, 9}

\bibitem[\protect\citeauthoryear{{Jiang}, {Blaes}, {Stone}  \& {Davis}}{{Jiang} et~al.}{2019}]{jia+19}
{Jiang} Y.-F.,  {Blaes} O.,  {Stone} J.~M.,   {Davis} S.~W.,  2019, \mn@doi [\apj] {10.3847/1538-4357/ab4a00}, \href {https://ui.adsabs.harvard.edu/abs/2019ApJ...885..144J} {885, 144}

\bibitem[\protect\citeauthoryear{{Kadowaki}, {De Gouveia Dal Pino}  \& {Stone}}{{Kadowaki} et~al.}{2018}]{kad+18}
{Kadowaki} L. H.~S.,  {De Gouveia Dal Pino} E.~M.,   {Stone} J.~M.,  2018, \mn@doi [\apj] {10.3847/1538-4357/aad4ff}, \href {https://ui.adsabs.harvard.edu/abs/2018ApJ...864...52K} {864, 52}

\bibitem[\protect\citeauthoryear{{Linares}}{{Linares}}{2014}]{2014Linares}
{Linares} M.,  2014, \mn@doi [\apj] {10.1088/0004-637X/795/1/72}, \href {https://ui.adsabs.harvard.edu/abs/2014ApJ...795...72L} {795, 72}

\bibitem[\protect\citeauthoryear{Linares et~al.,}{Linares et~al.}{2013}]{2014bLinares}
Linares M.,  et~al., 2013, \mn@doi [Monthly Notices of the Royal Astronomical Society] {10.1093/mnras/stt2167}, 438, 251

\bibitem[\protect\citeauthoryear{{Linares}, {De Marco}, {Wijnands}  \& {van der Klis}}{{Linares} et~al.}{2022}]{linares2022}
{Linares} M.,  {De Marco} B.,  {Wijnands} R.,   {van der Klis} M.,  2022, \mn@doi [\mnras] {10.1093/mnras/stac720}, \href {https://ui.adsabs.harvard.edu/abs/2022MNRAS.512.5269L} {512, 5269}

\bibitem[\protect\citeauthoryear{{Lipunov}}{{Lipunov}}{1992}]{1992Lipunov}
{Lipunov} V.~M.,  1992, {Astrophysics of Neutron Stars}.
Springer

\bibitem[\protect\citeauthoryear{{Lovelace}, {Romanova}  \& {Bisnovatyi-Kogan}}{{Lovelace} et~al.}{1999}]{lov+99}
{Lovelace} R.~V.~E.,  {Romanova} M.~M.,   {Bisnovatyi-Kogan} G.~S.,  1999, \mn@doi [\apj] {10.1086/306945}, \href {https://ui.adsabs.harvard.edu/abs/1999ApJ...514..368L} {514, 368}

\bibitem[\protect\citeauthoryear{{Menou}, {Perna}  \& {Hernquist}}{{Menou} et~al.}{2001}]{men+01}
{Menou} K.,  {Perna} R.,   {Hernquist} L.,  2001, \mn@doi [\apjl] {10.1086/320927}, \href {http://adsabs.harvard.edu/abs/2001ApJ...554L..63M} {554, L63}

\bibitem[\protect\citeauthoryear{{Michel}}{{Michel}}{1988}]{mic88}
{Michel} F.~C.,  1988, \mn@doi [\nat] {10.1038/333644a0}, \href {http://adsabs.harvard.edu/abs/1988Natur.333..644M} {333, 644}

\bibitem[\protect\citeauthoryear{{Michel} \& {Dessler}}{{Michel} \& {Dessler}}{1981}]{mic81}
{Michel} F.~C.,  {Dessler} A.~J.,  1981, \mn@doi [\apj] {10.1086/159511}, \href {http://adsabs.harvard.edu/abs/1981ApJ...251..654M} {251, 654}

\bibitem[\protect\citeauthoryear{{Michel} \& {Li}}{{Michel} \& {Li}}{1999}]{michel99}
{Michel} F.~C.,  {Li} H.,  1999, \mn@doi [\physrep] {10.1016/S0370-1573(99)00002-2}, \href {https://ui.adsabs.harvard.edu/abs/1999PhR...318..227M} {318, 227}

\bibitem[\protect\citeauthoryear{{Miller} \& {Stone}}{{Miller} \& {Stone}}{1997}]{miller1997}
{Miller} K.~A.,  {Stone} J.~M.,  1997, \mn@doi [\apj] {10.1086/304825}, \href {https://ui.adsabs.harvard.edu/abs/1997ApJ...489..890M} {489, 890}

\bibitem[\protect\citeauthoryear{{Misra} \& {Taam}}{{Misra} \& {Taam}}{2002}]{mis02}
{Misra} R.,  {Taam} R.~E.,  2002, \mn@doi [\apj] {10.1086/340764}, \href {https://ui.adsabs.harvard.edu/abs/2002ApJ...573..764M} {573, 764}

\bibitem[\protect\citeauthoryear{{Murguia-Berthier}, {Parfrey}, {Tchekhovskoy}  \& {Jacquemin-Ide}}{{Murguia-Berthier} et~al.}{2024}]{mur+24}
{Murguia-Berthier} A.,  {Parfrey} K.,  {Tchekhovskoy} A.,   {Jacquemin-Ide} J.,  2024, \mn@doi [\apjl] {10.3847/2041-8213/ad16eb}, \href {https://ui.adsabs.harvard.edu/abs/2024ApJ...961L..20M} {961, L20}

\bibitem[\protect\citeauthoryear{{Okazaki}, {Nagataki}, {Naito}, {Kawachi}, {Hayasaki}, {Owocki}  \& {Takata}}{{Okazaki} et~al.}{2011}]{okazaki2011}
{Okazaki} A.~T.,  {Nagataki} S.,  {Naito} T.,  {Kawachi} A.,  {Hayasaki} K.,  {Owocki} S.~P.,   {Takata} J.,  2011, \mn@doi [\pasj] {10.1093/pasj/63.4.893}, \href {https://ui.adsabs.harvard.edu/abs/2011PASJ...63..893O} {63, 893}

\bibitem[\protect\citeauthoryear{{{\"O}zs{\"u}kan}, {Ek{\c{s}}i}, {Hambaryan}, {Neuh{\"a}user}, {Hohle}, {Ginski}  \& {Werner}}{{{\"O}zs{\"u}kan} et~al.}{2014}]{ozs+14}
{{\"O}zs{\"u}kan} G.,  {Ek{\c{s}}i} K.~Y.,  {Hambaryan} V.,  {Neuh{\"a}user} R.,  {Hohle} M.~M.,  {Ginski} C.,   {Werner} K.,  2014, \mn@doi [\apj] {10.1088/0004-637X/796/1/46}, \href {https://ui.adsabs.harvard.edu/abs/2014ApJ...796...46O} {796, 46}

\bibitem[\protect\citeauthoryear{{Papitto} \& {de Martino}}{{Papitto} \& {de Martino}}{2022a}]{2022Papitto}
{Papitto} A.,  {de Martino} D.,  2022a, in {Bhattacharyya} S.,  {Papitto} A.,   {Bhattacharya} D.,  eds,  Astrophysics and Space Science Library Vol. 465, Astrophysics and Space Science Library. pp 157--200 (\mn@eprint {arXiv} {2010.09060}), \mn@doi{10.1007/978-3-030-85198-9_6}

\bibitem[\protect\citeauthoryear{{Papitto} \& {de Martino}}{{Papitto} \& {de Martino}}{2022b}]{papitto2022}
{Papitto} A.,  {de Martino} D.,  2022b, in {Bhattacharyya} S.,  {Papitto} A.,   {Bhattacharya} D.,  eds,  Astrophysics and Space Science Library Vol. 465, Astrophysics and Space Science Library. pp 157--200 (\mn@eprint {arXiv} {2010.09060}), \mn@doi{10.1007/978-3-030-85198-9_6}

\bibitem[\protect\citeauthoryear{{Parfrey} \& {Tchekhovskoy}}{{Parfrey} \& {Tchekhovskoy}}{2017}]{parfrey2017b}
{Parfrey} K.,  {Tchekhovskoy} A.,  2017, \mn@doi [\apjl] {10.3847/2041-8213/aa9c85}, \href {https://ui.adsabs.harvard.edu/abs/2017ApJ...851L..34P} {851, L34}

\bibitem[\protect\citeauthoryear{{Parfrey} \& {Tchekhovskoy}}{{Parfrey} \& {Tchekhovskoy}}{2024}]{par24}
{Parfrey} K.,  {Tchekhovskoy} A.,  2024, \mn@doi [\apj] {10.3847/1538-4357/ad737b}, \href {https://ui.adsabs.harvard.edu/abs/2024ApJ...975...57P} {975, 57}

\bibitem[\protect\citeauthoryear{{Parfrey}, {Spitkovsky}  \& {Beloborodov}}{{Parfrey} et~al.}{2017}]{parfrey2017}
{Parfrey} K.,  {Spitkovsky} A.,   {Beloborodov} A.~M.,  2017, \mn@doi [\mnras] {10.1093/mnras/stx950}, \href {https://ui.adsabs.harvard.edu/abs/2017MNRAS.469.3656P} {469, 3656}

\bibitem[\protect\citeauthoryear{{Philippov}, {Tchekhovskoy}  \& {Li}}{{Philippov} et~al.}{2014}]{philippov2014}
{Philippov} A.,  {Tchekhovskoy} A.,   {Li} J.~G.,  2014, \mn@doi [\mnras] {10.1093/mnras/stu591}, \href {https://ui.adsabs.harvard.edu/abs/2014MNRAS.441.1879P} {441, 1879}

\bibitem[\protect\citeauthoryear{{Price}}{{Price}}{2012}]{pri12}
{Price} D.~J.,  2012, \mn@doi [Journal of Computational Physics] {10.1016/j.jcp.2010.12.011}, \href {http://adsabs.harvard.edu/abs/2012JCoPh.231..759P} {231, 759}

\bibitem[\protect\citeauthoryear{{Price} et~al.,}{{Price} et~al.}{2018}]{price18}
{Price} D.~J.,  et~al., 2018, \mn@doi [\pasa] {10.1017/pasa.2018.25}, \href {https://ui.adsabs.harvard.edu/abs/2018PASA...35...31P} {35, e031}

\bibitem[\protect\citeauthoryear{{Pringle} \& {Rees}}{{Pringle} \& {Rees}}{1972}]{pringle1972}
{Pringle} J.~E.,  {Rees} M.~J.,  1972, \aap, \href {https://ui.adsabs.harvard.edu/abs/1972A&A....21....1P} {21, 1}

\bibitem[\protect\citeauthoryear{{Psaltis} \& {Chakrabarty}}{{Psaltis} \& {Chakrabarty}}{1999}]{psa99}
{Psaltis} D.,  {Chakrabarty} D.,  1999, \mn@doi [\apj] {10.1086/307525}, \href {https://ui.adsabs.harvard.edu/abs/1999ApJ...521..332P} {521, 332}

\bibitem[\protect\citeauthoryear{{Romanova} \& {Owocki}}{{Romanova} \& {Owocki}}{2015}]{Rom15}
{Romanova} M.~M.,  {Owocki} S.~P.,  2015, \mn@doi [\ssr] {10.1007/s11214-015-0200-9}, \href {https://ui.adsabs.harvard.edu/abs/2015SSRv..191..339R} {191, 339}

\bibitem[\protect\citeauthoryear{{Romanova}, {Ustyugova}, {Koldoba}  \& {Lovelace}}{{Romanova} et~al.}{2002}]{Rom+02}
{Romanova} M.~M.,  {Ustyugova} G.~V.,  {Koldoba} A.~V.,   {Lovelace} R.~V.~E.,  2002, \mn@doi [\apj] {10.1086/342464}, \href {https://ui.adsabs.harvard.edu/abs/2002ApJ...578..420R} {578, 420}

\bibitem[\protect\citeauthoryear{{Romanova}, {Toropina}, {Toropin}  \& {Lovelace}}{{Romanova} et~al.}{2003a}]{Rom+03pr}
{Romanova} M.~M.,  {Toropina} O.~D.,  {Toropin} Y.~M.,   {Lovelace} R.~V.~E.,  2003a, \mn@doi [\apj] {10.1086/373990}, \href {https://ui.adsabs.harvard.edu/abs/2003ApJ...588..400R} {588, 400}

\bibitem[\protect\citeauthoryear{{Romanova}, {Ustyugova}, {Koldoba}, {Wick}  \& {Lovelace}}{{Romanova} et~al.}{2003b}]{Rom+03inc}
{Romanova} M.~M.,  {Ustyugova} G.~V.,  {Koldoba} A.~V.,  {Wick} J.~V.,   {Lovelace} R.~V.~E.,  2003b, \mn@doi [\apj] {10.1086/377514}, \href {https://ui.adsabs.harvard.edu/abs/2003ApJ...595.1009R} {595, 1009}

\bibitem[\protect\citeauthoryear{{Romanova}, {Ustyugova}, {Koldoba}  \& {Lovelace}}{{Romanova} et~al.}{2004}]{Rom+04}
{Romanova} M.~M.,  {Ustyugova} G.~V.,  {Koldoba} A.~V.,   {Lovelace} R.~V.~E.,  2004, \mn@doi [\apjl] {10.1086/426586}, \href {https://ui.adsabs.harvard.edu/abs/2004ApJ...616L.151R} {616, L151}

\bibitem[\protect\citeauthoryear{{Romanova}, {Kulkarni}  \& {Lovelace}}{{Romanova} et~al.}{2008}]{Rom+08}
{Romanova} M.~M.,  {Kulkarni} A.~K.,   {Lovelace} R. V.~E.,  2008, \mn@doi [\apjl] {10.1086/527298}, \href {https://ui.adsabs.harvard.edu/abs/2008ApJ...673L.171R} {673, L171}

\bibitem[\protect\citeauthoryear{{Romanova}, {Blinova}, {Ustyugova}, {Koldoba}  \& {Lovelace}}{{Romanova} et~al.}{2018}]{Rom+18}
{Romanova} M.~M.,  {Blinova} A.~A.,  {Ustyugova} G.~V.,  {Koldoba} A.~V.,   {Lovelace} R.~V.~E.,  2018, \mn@doi [\na] {10.1016/j.newast.2018.01.011}, \href {https://ui.adsabs.harvard.edu/abs/2018NewA...62...94R} {62, 94}

\bibitem[\protect\citeauthoryear{{Romanova}, {Koldoba}, {Ustyugova}, {Blinova}, {Lai}  \& {Lovelace}}{{Romanova} et~al.}{2021}]{romanova2021}
{Romanova} M.~M.,  {Koldoba} A.~V.,  {Ustyugova} G.~V.,  {Blinova} A.~A.,  {Lai} D.,   {Lovelace} R.~V.~E.,  2021, \mn@doi [\mnras] {10.1093/mnras/stab1724}, \href {https://ui.adsabs.harvard.edu/abs/2021MNRAS.506..372R} {506, 372}

\bibitem[\protect\citeauthoryear{{Ronchi}, {Rea}, {Graber}  \& {Hurley-Walker}}{{Ronchi} et~al.}{2022}]{ron+22}
{Ronchi} M.,  {Rea} N.,  {Graber} V.,   {Hurley-Walker} N.,  2022, \mn@doi [\apj] {10.3847/1538-4357/ac7cec}, \href {https://ui.adsabs.harvard.edu/abs/2022ApJ...934..184R} {934, 184}

\bibitem[\protect\citeauthoryear{{Rosswog}}{{Rosswog}}{2010}]{Rosswog2010}
{Rosswog} S.,  2010, \mn@doi [Journal of Computational Physics] {10.1016/j.jcp.2010.08.002}, \href {https://ui.adsabs.harvard.edu/abs/2010JCoPh.229.8591R} {229, 8591}

\bibitem[\protect\citeauthoryear{{Shahbaz}, {Linares}, {Rodr{\'\i}guez-Gil}  \& {Casares}}{{Shahbaz} et~al.}{2019}]{shahbaz2019}
{Shahbaz} T.,  {Linares} M.,  {Rodr{\'\i}guez-Gil} P.,   {Casares} J.,  2019, \mn@doi [\mnras] {10.1093/mnras/stz1652}, \href {https://ui.adsabs.harvard.edu/abs/2019MNRAS.488..198S} {488, 198}

\bibitem[\protect\citeauthoryear{{Shakura} \& {Sunyaev}}{{Shakura} \& {Sunyaev}}{1973}]{Shakura}
{Shakura} N.~I.,  {Sunyaev} R.~A.,  1973, \aap, \href {https://ui.adsabs.harvard.edu/abs/1973A&A....24..337S} {24, 337}

\bibitem[\protect\citeauthoryear{{Shvartsman}}{{Shvartsman}}{1970}]{shvartsman1970}
{Shvartsman} V.~F.,  1970, \sovast, \href {https://ui.adsabs.harvard.edu/abs/1970SvA....14..527S} {14, 527}

\bibitem[\protect\citeauthoryear{{Shvartsman}}{{Shvartsman}}{1971}]{shvartsman1971}
{Shvartsman} V.~F.,  1971, \sovast, \href {https://ui.adsabs.harvard.edu/abs/1971SvA....15..342S} {15, 342}

\bibitem[\protect\citeauthoryear{{Stappers} et~al.,}{{Stappers} et~al.}{2014}]{stappers2014}
{Stappers} B.~W.,  et~al., 2014, \mn@doi [\apj] {10.1088/0004-637X/790/1/39}, \href {https://ui.adsabs.harvard.edu/abs/2014ApJ...790...39S} {790, 39}

\bibitem[\protect\citeauthoryear{{Takasao}, {Tomida}, {Iwasaki}  \& {Suzuki}}{{Takasao} et~al.}{2018}]{tak+18}
{Takasao} S.,  {Tomida} K.,  {Iwasaki} K.,   {Suzuki} T.~K.,  2018, \mn@doi [\apj] {10.3847/1538-4357/aab5b3}, \href {https://ui.adsabs.harvard.edu/abs/2018ApJ...857....4T} {857, 4}

\bibitem[\protect\citeauthoryear{{Tricco}}{{Tricco}}{2023}]{tricco2023}
{Tricco} T.~S.,  2023, \mn@doi [Frontiers in Astronomy and Space Sciences] {10.3389/fspas.2023.1288219}, \href {https://ui.adsabs.harvard.edu/abs/2023FrASS..1088219T} {10, 1288219}

\bibitem[\protect\citeauthoryear{{Tricco}, {Price}  \& {Bate}}{{Tricco} et~al.}{2016}]{tricco2016}
{Tricco} T.~S.,  {Price} D.~J.,   {Bate} M.~R.,  2016, \mn@doi [Journal of Computational Physics] {10.1016/j.jcp.2016.06.053}, \href {https://ui.adsabs.harvard.edu/abs/2016JCoPh.322..326T} {322, 326}

\bibitem[\protect\citeauthoryear{{Veledina}, {N{\"a}ttil{\"a}}  \& {Beloborodov}}{{Veledina} et~al.}{2019}]{veledina2019}
{Veledina} A.,  {N{\"a}ttil{\"a}} J.,   {Beloborodov} A.~M.,  2019, \mn@doi [\apj] {10.3847/1538-4357/ab44c6}, \href {https://ui.adsabs.harvard.edu/abs/2019ApJ...884..144V} {884, 144}

\bibitem[\protect\citeauthoryear{{Wang}}{{Wang}}{1996}]{Wang96}
{Wang} Y.~M.,  1996, \mn@doi [\apjl] {10.1086/310150}, \href {https://ui.adsabs.harvard.edu/abs/1996ApJ...465L.111W} {465, L111}

\bibitem[\protect\citeauthoryear{{Wang}, {Chakrabarty}  \& {Kaplan}}{{Wang} et~al.}{2006}]{wan+06}
{Wang} Z.,  {Chakrabarty} D.,   {Kaplan} D.~L.,  2006, \mn@doi [\nat] {10.1038/nature04669}, \href {http://adsabs.harvard.edu/abs/2006Natur.440..772W} {440, 772}

\bibitem[\protect\citeauthoryear{{Wijnands} \& {Degenaar}}{{Wijnands} \& {Degenaar}}{2013}]{Wijnands13}
{Wijnands} R.,  {Degenaar} N.,  2013, \mn@doi [\mnras] {10.1093/mnras/stt1119}, \href {https://ui.adsabs.harvard.edu/abs/2013MNRAS.434.1599W} {434, 1599}

\bibitem[\protect\citeauthoryear{{Wijnands}, {Degenaar}, {Armas Padilla}, {Altamirano}, {Cavecchi}, {Linares}, {Bahramian}  \& {Heinke}}{{Wijnands} et~al.}{2015}]{Wijnands15}
{Wijnands} R.,  {Degenaar} N.,  {Armas Padilla} M.,  {Altamirano} D.,  {Cavecchi} Y.,  {Linares} M.,  {Bahramian} A.,   {Heinke} C.~O.,  2015, \mn@doi [\mnras] {10.1093/mnras/stv1974}, \href {https://ui.adsabs.harvard.edu/abs/2015MNRAS.454.1371W} {454, 1371}

\bibitem[\protect\citeauthoryear{{Wissing} \& {Shen}}{{Wissing} \& {Shen}}{2020}]{wissing2020}
{Wissing} R.,  {Shen} S.,  2020, \mn@doi [\aap] {10.1051/0004-6361/201936739}, \href {https://ui.adsabs.harvard.edu/abs/2020A&A...638A.140W} {638, A140}

\bibitem[\protect\citeauthoryear{{Zanni} \& {Ferreira}}{{Zanni} \& {Ferreira}}{2009}]{zanni2009}
{Zanni} C.,  {Ferreira} J.,  2009, \mn@doi [\aap] {10.1051/0004-6361/200912879}, \href {https://ui.adsabs.harvard.edu/abs/2009A&A...508.1117Z} {508, 1117}

\bibitem[\protect\citeauthoryear{{Zanni} \& {Ferreira}}{{Zanni} \& {Ferreira}}{2013}]{zanni2013}
{Zanni} C.,  {Ferreira} J.,  2013, \mn@doi [\aap] {10.1051/0004-6361/201220168}, \href {https://ui.adsabs.harvard.edu/abs/2013A&A...550A..99Z} {550, A99}

\bibitem[\protect\citeauthoryear{{{\c C}al{\i}{\c s}kan}, {Ertan}, {Alpar}, {Tr{\"u}mper}  \& {Kylafis}}{{{\c C}al{\i}{\c s}kan} et~al.}{2013}]{cal+13}
{{\c C}al{\i}{\c s}kan} {\c S}.,  {Ertan} {\"U}.,  {Alpar} M.~A.,  {Tr{\"u}mper} J.~E.,   {Kylafis} N.~D.,  2013, \mn@doi [\mnras] {10.1093/mnras/stt234}, \href {http://adsabs.harvard.edu/abs/2013MNRAS.431.1136C} {431, 1136}

\bibitem[\protect\citeauthoryear{{{\c{C}}{\i}k{\i}nto{\u{g}}lu}, {Ek{\c{s}}i}  \& {Rezzolla}}{{{\c{C}}{\i}k{\i}nto{\u{g}}lu} et~al.}{2022}]{cik+22}
{{\c{C}}{\i}k{\i}nto{\u{g}}lu} S.,  {Ek{\c{s}}i} K.~Y.,   {Rezzolla} L.,  2022, \mn@doi [\mnras] {10.1093/mnras/stac2510}, \href {https://ui.adsabs.harvard.edu/abs/2022MNRAS.517.3212C} {517, 3212}

\makeatother
\end{thebibliography}




\appendix

\section{Building balanced axisymmetric discs with SPH}
\label{AppA}

To relax the disc, we make use of the technique described in \cite{2020Garcia}. First, SPH particles are radially distributed according to the density profile $\rho(r)$, and randomly distributed in the vertical z direction, but with a weight such that $\rho(r,z)$ exponentially decays with the altitude at the coordinate of the disc $r$, as required by the thin disc solution. The angular velocity of the particle $a$ of the disc is first set to
\begin{equation}
    \omega_a=\frac{\omega_c}{\left(1+\frac{r_a^2}{r_c^2}\right)^m}\ ,
\end{equation}
with $m=3/4$ (Keplerian disc), $r_c=2\times 10^6$ cm and $\omega_c$ is a time-dependent constant determined to enforce the conservation of the total angular momentum, $J_{\rm disc}$, of the disc \citep{2020Garcia} in each iteration step. In each integration step, particles are displaced a distance proportional to the value of the particle acceleration and to its smoothing length $h$. This procedure ensures that all regions of the disc achieve equilibrium, even those regions of the disc where the characteristic relaxation time is larger. The coordinates $(r, z)$ of the particles are updated with the following prescription,
\begin{equation}
    (r,z)=(r_0, z_0)+ \eta h \left(\frac{\vert a_r\vert}{\sqrt{f^2_{p_r} + f^2_{g_r}+ f^2_c}} u_r, \frac{\vert a_z\vert}{\sqrt{f^2_{p_z}+ f^2_{g_z}}} u_z \right)\ ,  
    \label{relax_displacement}
\end{equation}
where $f_p, f_g, f_c$ symbolises the pressure, gravity and centripetal forces, respectively, and $(u_r, u_z)$ is the unit vector pointing along the acceleration vector $(a_r, a_z)$. The parameter $\eta\simeq 2\cdot 10^{-4}\times (r/R_{\rm in})^n$, where $R_{\rm in}$ is the position of the inner radius of the disc and $n\simeq 0.6$, controls the amount of displacement so that it is larger in the regions of the disc with $r \gg R_{\rm in}$. In this way, the initial 2D distribution of SPH particles settles in a balanced disc structure which preserves the total mass and total angular momentum obtained with the analytical thin disc approach described in \cite{frank02}. 



\bsp	
\label{lastpage}
\end{document}